\documentclass[pagebackref=true,11pt]{amsart}
\usepackage{hyperref}
\hypersetup{
    colorlinks,
    citecolor=blue,
    filecolor=blue,
    linkcolor=blue,
    urlcolor=blue 
}
\usepackage{orcidlink}
\usepackage[numbers]{natbib}
\usepackage[margin=1 in]{geometry}
\usepackage{graphicx}%
\usepackage{multirow}%
\usepackage{amsfonts}%
\usepackage{textcomp}%
\usepackage{manyfoot}%
\usepackage{CJKutf8}
\usepackage{booktabs}%
\usepackage{algorithm}%
\usepackage{algorithmic}%
\usepackage{listings}%
\usepackage{xcolor}
\usepackage{tikz}
\usepackage{tikz-cd}
\usepackage{tikz-3dplot}
\usepackage{amsmath,mathrsfs,amssymb,amsthm,amscd}
\usepackage{mathtools}
\usepackage{url}
\usepackage{natbib}
\usepackage{epigraph}
\usepackage[all]{xy}
\usepackage{wasysym}
\newtheorem*{rep@theorem}{\rep@title}
\newcommand{\newreptheorem}[2]{%
\newenvironment{rep#1}[1]{%
 \def\rep@title{#2 \ref{##1} (restated)}%
 \begin{rep@theorem}}%
 {\end{rep@theorem}}}
\makeatother

\newtheorem{theorem}{Theorem}[section]
\newtheorem{lemma}[theorem]{Lemma}
\newtheorem{proposition}[theorem]{Proposition}

\newreptheorem{theorem}{Theorem}

\theoremstyle{definition}
\newtheorem{definition}[theorem]{Definition}
\newtheorem*{question}{Question}

\theoremstyle{remark}
\newtheorem*{remark}{Remark}

\numberwithin{equation}{section}

\begin{document}


\title[The factor $P_1(T)$ of the zeta function]{Complexity of counting points on curves, and the factor $P_1(T)$ of the zeta function of surfaces}


\author{Diptajit Roy}
\address{Department of Computer Science \& Engineering, IIT Kanpur, India} 
\curraddr{}
\email{diptajit@cse.iitk.ac.in}
\thanks{}

\author{Nitin Saxena \orcidlink{0000-0001-6931-898X}}
\address{}
\curraddr{\url{https://www.cse.iitk.ac.in/users/nitin/}}
\email{nitin@cse.iitk.ac.in}
\thanks{}

\author{Madhavan Venkatesh}

\curraddr{}
\email{madhavan@cse.iitk.ac.in}



\date{}

\dedicatory{}
\begin{abstract}
This article concerns the computational complexity of a fundamental 
problem in number theory: counting points on curves and surfaces over finite 
fields. There is no subexponential-time algorithm known and it is unclear if it can be $\mathrm{NP}$-hard.

Given a curve (say,  $f(x,y)=0$ of degree $d$ over field $\mathbb{F}_{q}$), we present the first efficient Arthur-Merlin protocol to certify its point-count, its Jacobian group structure, and its Hasse-Weil zeta function. We place this problem in $\mathrm{AM}\cap\mathrm{coAM}$, while the previous best was BQP (Kedlaya'06). We extend this result to a smooth projective \textit{surface} (say, dimension $2$ in $\mathbb{P}^4$ and degree $D$) to certify the factor $P_{1}(T)$, corresponding to the \textit{first} Betti number, of the zeta function; the previous best was P$^{\#\rm P}$ by using the counting oracle. Famously, the complex reciprocal roots of $P_{1}(T)$ have norm $\sqrt{q}$ (Deligne's proof of the Weil-Riemann Hypothesis, 1974), and it tells us all about the Picard variety of the surface. We give the first algorithm to compute $P_{1}(T)$ that is poly($\log q$)-time if the degree $D$ of the input surface is \textit{fixed}; and in quantum poly($D\log q$)-time in general.

Our technique in the curve case, is to sample  hash functions using the Weil and Riemann-Roch bounds, to certify the group order of its Jacobian. For \textit{higher} dimension varieties, we first reduce to the case of a surface, which is fibred as a Lefschetz pencil of hyperplane sections over $\mathbb{P}^{1}$. The formalism of vanishing cycles, and the inherent \textit{big monodromy}, enable us to prove an effective version of Deligne's `theoreme du pgcd' using the hard-Lefschetz theorem and an equidistribution result due to Katz. These reduce our investigations to that of computing the zeta function of a curve, defined over a finite field extension $\mathbb{F}_{Q}/\mathbb{F}_{q}$ of poly-bounded degree. This explicitization of the theory yields the first nontrivial upper bounds on the computational complexity.

\end{abstract}

\maketitle
\vspace{-1.1cm}
\setcounter{tocdepth}{1}
\tableofcontents

\vspace{-1.3cm}
\section{Introduction}\label{sec1}
Since antiquity mathematicians have studied `simple' equations to find, and count, the roots; unearthing powerful theories. A classical family is the projective curve $C$ (in $\mathbb{P}^{2}$): $a_0x_0^{d}+a_1x_1^{d}+a_2x_2^{d} \equiv0 \bmod{p}$, and the projective surface $S$ (in $\mathbb{P}^{3}$): $a_0x_0^{d}+a_1x_1^{d}+a_2x_2^{d}+a_3x_3^{d} \equiv0 \bmod{p}$, for a prime $p$, and numbers $a_i$'s, $d$. One would like to count the roots, denoted $|C(\mathbb{F}_p)|$ resp.~$|S(\mathbb{F}_p)|$, in time polynomial in $\log p$ and $d$. We can trivially estimate the counts to be $p$ resp.~$p^2$, but how good are these estimates? This has been studied, for various cases, at least since the times of Gau\ss~(1800s) \cite{pierpont1903gauss}, Jacobi \cite{jacobi1884cgj}, Lebesgue, Hardy \& Littlewood, Davenport \& Hasse; till the modern formulation of Weil-Riemann hypothesis of a zeta function was given by Weil \cite{weil, WeilA}. It uses the topological and geometric properties of a variety to reflect on its arithmetic properties. 

Specifically, over characteristic zero, one can associate to a smooth variety its singular and de Rham cohomology groups. In this setting, there have been algorithmic results on computing these topological invariants e.g., the number of irreducible components in \cite{buerg}, and more general cohomology computations using real algebraic geometry \cite{basu, scheib} and algebraic de Rham cohomology \cite{oaku, scheib2}.

We are interested in an arithmetic analogue of this line of work.
This study has found numerous applications in modern computing; especially motivated by the example of curves \cite{cohen2005handbook} and surfaces \cite{berardini}. In particular, the genus of a curve and its number of rational points can be read off from its zeta function, and selecting a curve with optimised such parameters is a natural question that crops up in the theory of algebraic-geometric (AG) codes \cite{agc}. For a higher dimensional variety, the first cohomology encodes information about its \textit{Picard} variety, a natural abelian variety parametrising codimension one subvarieties modulo an equivalence relation; that plays the analogue of the Jacobian of a curve.

 In this work we will clarify the complexity of the curve case in a significant way, and we will take the first steps in the surface case. The topological invariants involved in the latter are much harder as they demand the most abstruse cohomology theory \cite{fk}. In particular, we provide the first explicit, computational results on the Picard variety of higher dimensional varieties.

Let $X_{0}$ be a smooth, projective variety of dimension $n$ over the finite field $\mathbb{F}_{q}$ of characteristic $p>0$. Denote by $X$ the base-change to the algebraic closure $\overline{\mathbb{F}}_{q}$. To encode the number of its points over all finite field extensions, the zeta function of $X$ is defined as 
$$
Z(X/\mathbb{F}_{q}, T)\, :=\,\text{exp}\left(\sum_{j=1}^{\infty}\# X(\mathbb{F}_{q^j})\frac{T^j}{j}\right) \;\in\; \mathbb{Z}[[T]] \,.
$$
It is an exponential of the generating function of the {\em point-counts}. The result is seen as a formal power series in $T$. 
Let $\ell$ be a prime distinct from $p$. By the foundational work of Grothendieck et al.\cite{sgav} on $\ell$-adic cohomology, it is known that the zeta function can be written as a rational function:
\begin{equation}
\label{eq:zeta}
Z(X/\mathbb{F}_q,T) \,=\, \frac{P_1(T)P_3(T)\cdots P_{2n-1}(T)}{P_0(T)P_2(T)\cdots P_{2n}(T)}\in \mathbb{Q}(T) \,,
\end{equation}
where $P_{i}(T)=\mathrm{det}(1-TF_{q}^{\star} \ \vert \ \mathrm{H}^{i}(X, \mathbb{Q}_{\ell}))$ is the characteristic polynomial\footnote{or \textit{reversed} characteristic polynomial, according to another convention} of the map $F_{q}^{\star}$ induced on the cohomology by the geometric Frobenius. Further, the zeta function satisfies the functional equation
$$
Z(X/\mathbb{F}_{q},1/(q^{n}T) )\;=\;\pm q^{n\cdot\chi/2}\cdot T^{\chi}\cdot Z(X/\mathbb{F}_{q}, T),
$$
where $\chi=\sum_{i=0}^{2n}(-1)^i\cdot\dim \mathrm{H}^{i}(X, \mathbb{Q}_{\ell})$ is the $\ell$-adic Euler-Poincar\'e characteristic of $X$. Denote $\beta_{i}:=\dim \mathrm{H}^{i}(X, \mathbb{Q}_{\ell})$, also called the $i$th Betti number.
As a result of Deligne's proof \cite{Weili} of the Weil-Riemann hypothesis, we have
$P_{i}(T)=\prod_{j=1}^{\beta_{i}}(1-\alpha_{i,j}T)\in \mathbb{Z}[T]$, with $\alpha_{i,j}$ being algebraic numbers such that $\vert \iota(\alpha_{i,j})\vert=q^{i/2}$ for any embedding $\iota:\mathbb{Q}(\alpha_{i,j})\rightarrow \mathbb{C}$. In particular, it follows that the $P_{i}(T)$ are independent of $\ell$. 

The complexity of computing the zeta function of a variety over a finite field is a natural question, being the generalisation of the ancient problem of counting the number of congruent solutions of a given polynomial equation modulo a prime $p$. Let $X\subset \mathbb{P}^{N}$ be a smooth, projective variety of dimension $n$ and degree $D$, presented as the zero set of homogeneous polynomials $f_{1}, \dots , f_{m}$ each of total degree $\leq d$. The dimension $n$ of the variety (and that of its embedding space, $N$) is considered {\em fixed}. This is because the Betti numbers of a variety, and hence also the degree of its zeta function, are {\em exponential} in $N$. 


So, in practice, one seeks algorithms to compute $Z(X/\mathbb{F}_{q}, T)$ efficiently in only two parameters, namely $\log q$ and $D$. Such an algorithm which is polynomial-time in both is unknown, despite being a well-studied problem in the intersection of mathematics and computer science. A special case of a question of Serre \cite[Preface]{serre2016lectures} asks the following (paraphrased), which fundamentally motivates our work.

\begin{question}[Compute]
    Let $\mathcal{X}_{0}$ be a (fixed) smooth, projective variety over $\mathbb{Q}$. Is there an algorithm which, given a prime $p$ of good reduction of $\mathcal{X}_{0}$; computes the point-count of the reduction, $\#X(\mathbb{F}_{p})$, in time polynomial in $\log p$?
\end{question}
 We obtain the first polynomial-time (in $\log q$) algorithm to compute $P_{1}(T)$ for smooth varieties (of dimension $\geq 2$) of fixed degree $D$, extending a line of work that goes back to elliptic curves \cite{schoof} and abelian varieties \cite{Pila}. Consequently, for a surface $X$, we can now compute all $P_i$'s except $P_2(T)$; thus, computing $Z(X/\mathbb{F}_{q}, T)\cdot P_2(T)$.

One notices that even for the simple-to-present hyperelliptic curves, $y^2=f(x) \bmod p$, that are quite useful in cryptography \cite{cohen2005handbook}, there is no fast algorithm known to compute the zeta function, in time polynomial in both $\log p$ and $\deg(f)$. So, one wonders if an `easier' {\em verification} question (see \cite[Question 15]{aim}) should be answered first: 

\begin{question}[Certify]
Given a variety $X$, a rational function $Q(T)$ and some `data', is there a polynomial-time algorithm to \textit{verify} that $Q(T)$ is the zeta function of $X$? In other words, is the zeta function computation problem in $\mathrm{NP}$, or in $\mathrm{coNP}$? More generally, given input polynomials $Q_i(T)\in\mathbb{Z}[T]$,  for all $i$, is {\em verifying} $$
Q_i(T)\stackrel{?}{=}P_i(T)
\quad\text{ in }\quad  \mathrm{NP}\cap \mathrm{coNP} \,? $$
\end{question}
In this work, we take a major step towards answering the above question about verifying the zeta function. Unfortunately, our protocol does not translate into a practical algorithm. But, we do show that the problem of computing zeta function of a smooth projective curve (with $D, \log q$ variables) is {\em unlikely} to be NP-hard, or has `intermediate' complexity (in the sense of \cite[\S8.2.4]{AB}).


\smallskip
Further, generalising work of Kedlaya \cite{kedlaya2006quantum} (which was restricted to curves), we obtain the first {\em quantum} algorithm for computing $P_{1}(T)$ that is polynomial-time in $\log q$ and $D$.

\subsection{Prior work}
\label{sec 1.1}
It is possible to interpret (\ref{eq:zeta}) via a trace formula in a suitable \textit{Weil cohomology theory}. Examples include $\ell$-adic cohomology, for primes $\ell$ distinct from the characteristic, developed by Grothendieck \cite{sgav}; and rigid cohomology, an extension of crystalline cohomology due to Berthelot \cite{berthelot}.
In general, algorithms for computing the zeta function can be classified broadly into two distinct families, $\ell$-adic resp.~ $p$-adic, usually based on the nature of the cohomology theory being employed. The progenitor of the $\ell$-adic class of algorithms is the work of Schoof \cite{schoof}, who gave an algorithm to compute the zeta function of an elliptic curve over $\mathbb{F}_{q}$ with complexity polynomial in $\log q$. This method was generalised by Pila \cite{Pila} to curves (of genus $g$), and abelian varieties, with improvements for some special cases due to Huang-Ierardi \cite{huaier} and Adleman-Huang \cite{AdleHwang}. The complexity of these algorithms, while polynomial in $\log q$ is exponential or worse in $g$. A common theme is the  realisation of the \'etale cohomology $\mathrm{H}^{1}(X, \mu_{\ell})$ as the $\ell$-torsion $\mathrm{Pic}^{0}(X)[\ell]$ in the Picard scheme. This has, so far, limited their application to varieties where this realisation can be made explicit, namely curves and abelian varieties. There has been work showing the computability of \'etale cohomology in higher degrees as well \cite{mad}, but it has not proven amenable to complexity analysis yet. 

On the other hand, $p$-adic methods encompass a more diverse range of algorithms. Some early examples are Satoh's algorithm for elliptic curves \cite{satoh} using canonical lifts and Kedlaya's algorithm for hyperelliptic curves \cite{KedlayapAdic}  using Monsky-Washnitzer cohomology (and extensions thereof \cite{Denef,cdv}). Lauder-Wan \cite{LW}, inspired by work of Dwork on the rationality of the zeta function \cite{dworkPadic}, proposed a more general algorithm capable of handling arbitrary varieties. Lauder \cite{lauder} also developed an algorithm for hypersurfaces based on $p$-adic deformation theory. More recently, there is the `non-cohomological' work of Harvey \cite{harvey}, who devised an algorithm based on a novel trace formula. The complexity of these algorithms, while polynomial in the degree $D$ of the variety, is exponential in $\log p$. A common theme is that they involve a $p$-adic lift of the Frobenius, which necessitates working with $O(p)$ monomials in the basis for the respective $p$-adic cohomology theory.

\subsection{A detour to basic complexity notions \cite{AB} }
 Since the zeta function is defined via an infinite sum of point-counts, the problem of computing $P_1(T)$ of a smooth projective variety could potentially be {\em un}computable! A lot of work has been done to pinpoint the {\em complexity} of this problem \cite{mad}; but a complete solution is unknown even in the case of smooth projective curves. 
 

This paper is motivated by the class of {\em Interactive Protocol}, where the verification algorithm (called {\em Arthur}) is allowed to have a number of interactions with the oracle ({\em Merlin}). In the Arthur-Merlin class, denoted by $\mathrm{AM}$, we assume that Arthur has access to Merlin only once throughout the computation.  Arthur is allowed to use randomisation in the verification algorithm (thus, it is like a randomised $\mathrm{NP}$ protocol). Problems lying in $\mathrm{AM} \cap \mathrm{coAM}$ class are `unlikely' to be NP-hard (else, the polynomial-hierarchy collapses, see \cite{bop}); optimistically, we can even conjecture them to have quasipolynomial-time algorithms.  Many famous problems are known to be in $\mathrm{AM}\cap\mathrm{coAM}$ -- e.g.~integer factoring, discrete logarithm, graph isomorphism, algebra isomorphism, and algebraic dependence (see \cite{kayal2006complexity, guo2019algebraic} and the references therein). A major byproduct of this work is to conclude that {\em computing $P_1(T)$ is unlikely to be NP-hard}, as we show it to be in $\mathrm{AM}\cap\mathrm{coAM}$.

Another popular complexity class is that of {\em quantum} polynomial-time, denoted BQP. It is not clear how it compares with the complexity classes we defined earlier, except the trivial comparison of $\mathrm{BPP} \subseteq \mathrm{BQP}$. Many famous problems are known to be in $\mathrm{BQP}$ --- e.g.~integer factoring, discrete logarithm, zeta function of curves, and the hidden-subgroup problem of abelian groups (see \cite{kedlaya2006quantum, nielsen2001quantum}). It is unknown if there is any NP-hard problem contained in BQP, or if BQP $\subseteq \mathrm{NP}\cup \mathrm{coNP}$. Similarly, BQP and AM are (currently) incomparable classes. Both of them are solvable using the {\em counting class} \#P as an oracle (e.g.~the problem of counting satisfying assignments).

\subsection{Main results: Certify or Compute}
\label{sec 1.2}

\hfill

\smallskip \noindent {\bf Certification.}
For a smooth, projective, geometrically irreducible {\em curve} $C\subset \mathbb{P}^{N}$ of genus $g>0$, the zeta function has the form
$$Z(C/\mathbb{F}_q, T)=\frac{P_{1}(C/\mathbb{F}_{q}, T)}{(1-T)(1-qT)},$$ where $P_1(C/\mathbb{F}_{q}, T)\in \mathbb{Z}[T]$ is of degree $2g$, with constant term $1$. 
Somewhat surprisingly, we will not only verify $P_1(C/\mathbb{F}_{q}, T)$ but also the abelian group structure of the Jacobian variety over the base field. It addresses a question of Kedlaya \cite[\S 9]{kedlaya2006quantum} on verifying the order of the Jacobian as a black-box group.

\begin{theorem}[Zeta \& Jacobian]
\label{Cohomology of curves}
    Given an input polynomial $P(T)\in \mathbb{Z}[T]$, deciding whether $P(T)$ is the numerator polynomial of the zeta function of the smooth, projective curve $C$, given as above (with  variable $g\log q$), is in $\mathrm{AM}\cap \mathrm{coAM}$.   
Moreover, given a finite Abelian group $G$ (via additive generators), the verification problem  
$$
G\stackrel{?}{\simeq} \mathrm{Jac}(C)(\mathbb{F}_q) \quad
\text { is in } \quad \mathrm{AM}\cap \mathrm{coAM} \,. $$    
\end{theorem}

 The above protocol reduces to the verification of a few group orders $\mathcal{N}_{j}:= \#\mathrm{Jac}(C)(\mathbb{F}_{q^{j}})$ of the Jacobian of $C$, which entails the verification of independence for a set of generators. The well-known ``mod-$\ell$ pairing''-based arguments do not give a protocol immediately; as, the order $\ell\mid\mathcal{N}_j$ of a generator  can be very large. In which case, it can require an exponential degree extension $\mathbb{F}_{Q}/\mathbb{F}_{q}$ for the Tate pairing to be non-degenerate on $\mathrm{Jac}(\mathbb{F}_{Q})[\ell]$ \cite{FreyRuck}. 

 \hfill

 We now assume the input is a smooth, projective {\em variety} $X_{0}\subset \mathbb{P}^{N}$ of dimension\footnote{the dimension of the embedding space, $N$, is considered to be fixed (say, $N=2, 3$ for $n=1, 2$ respectively), as we are primarily interested in {\em curves} and {\em surfaces} in this work. The degree $D$ and field size $q$ are allowed to vary.} $n\geq 1$ and degree $D$, over the finite field $\mathbb{F}_{q}$, presented as a system of $m$ homogeneous polynomials $f_1,\ldots,f_m$ of degree $\leq d$ \footnote{the complexity can be measured in $D$ or $d$, as for $N$ fixed, each is bounded by a polynomial function in the other.}. Further, we assume $X_{0}$ is obtained via good reduction of a smooth projective variety $\mathcal{X}_{0}$ over a number field $K$ at a prime $\mathfrak{p}$. As we are interested in the regime of varying the characteristic,  we assume accordingly that we have a globally defined {\em smooth} model over characteristic zero (i.e., a number field). Write $X:=X_{0}\times \overline{\mathbb{F}}_{q}$. We have the following certification result.
\begin{theorem}[Certify $P_1$]
\label{Theorem 1}
    Given $Q_1(T)\in \mathbb{Z}[T]$, deciding whether $Q_1(T)=P_{1}(X/\mathbb{F}_{q}, T)$, for $X_{0}$ given as above (with variable $D\log q$), is in $\mathrm{AM}\cap \mathrm{coAM}$. 
\end{theorem}
    
    The technical heart of the results in this work lies in the proof of Theorem~\ref{effgcd}, an effective version of Deligne's `th\'eor\`eme du pgcd' (from the celebrated work \cite{Weilii}). This allows us to reduce the computation of $P_{1}(T)$ \footnote{we write $P_{1}(X/ \mathbb{F}_{q}, T)$ to specify the $q$-power Frobenius.} for $X$ to the computation of the zeta function of smooth curves obtained by taking successive hyperplane sections of $X$, while the result for curves is proved in Theorem~\ref{Cohomology of curves}.

\smallskip\noindent {\bf Algorithmic results.} 
We also give the first quantum polynomial-time algorithm (allowing the degree $D$ to vary) to compute $P_{1}(X/\mathbb{F}_{q}, T)$, by applying Kedlaya's algorithm \cite{kedlaya2006quantum} for the curve case.
\begin{theorem}[Quantum]
\label{Theorem 3}
    There exists a quantum algorithm that computes $P_{1}(X/\mathbb{F}_{q}, T)$ in time polynomial in $D \log q$, for any $X_{0}$ given as above. 

\end{theorem}

For varieties of constant degree $D$, by our reduction to the case of curves and applying work of Pila \cite{Pila} and Huang-Ierardi \cite{huaier}, we have the following. 
\begin{theorem}[Fixed $D$]
\label{Theorem 2}
There exists a classical randomised algorithm that, given $X_{0}$ as above of {\em fixed} degree $D$, computes $P_{1}(X/\mathbb{F}_{q}, T)$ in time polynomial in $\log q$. 
\end{theorem}
 A major obstacle to computing the above was the lack of a concise and explicit representation, in general, for the \'etale cohomology group $\mathrm{H}^{1}(X, \mu_{\ell})$; despite it being known to be isomorphic to the $\ell$-torsion in the Picard scheme of $X$. A priori, elements therein are a formal sum of codimension-1 subvarieties (modulo an equivalence relation), and it is uncertain how one may directly produce $\ell$-torsion elements due to the highly {\em non-explicit} nature of the group law. There has been a strategy laid out by Levrat  \cite[IV.3.5, VI.4]{levrat} for surfaces, under some strongly restrictive hypotheses; but the general-case complexity is either unclear or exponential-time (see also \cite[\S 5]{leve}). 

\begin{remark}
In Theorems~\ref{Theorem 3} and~\ref{Theorem 2}, the stated runtimes are bounded by polynomial functions of the data as claimed, whose degree and coefficients are independent of  $D\cdot \log q$ for Theorem~\ref{Theorem 3} and $\log q$ for Theorem~\ref{Theorem 2}.
\end{remark}





\subsection{New techniques and proof ideas}
Certifying the zeta function of a smooth curve $C/\mathbb{F}_{q}$ of genus $g$ boils down to a certification of the group orders $\#\mathrm{Jac}(C)(\mathbb{F}_{q^{j}})$ for $1\leq j\leq 2g$. The addition law on the Jacobian can be made explicit (after reducing to a plane model) by an effective Riemann-Roch algorithm (Algorithm \ref{addtwopoints}). Utilising the additive structure of $\mathrm{Jac}(C)(\mathbb{F}_{q})\simeq \mathbb{Z}/n_{1}\times \ldots \times \mathbb{Z}/{n_{r}}$ (with $n_{i}\vert n_{i+1}$ and $r\leq 2g$) as an abelian group, it suffices to certify that a candidate generating-set of divisors $(D_{i})_{1\leq i\leq r}$, with each $D_{i}$ of order $n_{i}$, actually generates the full group. Using the Weil bound for the size of the Jacobian, we are able to certify, with high probability, the `independence' of the divisors $D_{i}$ (Algorithm \ref{algzetacurve}). This is done by random sampling in a family of hash functions--- a classical technique that originated in the famous protocol of Goldwasser-Sipser \cite{gs} to certify the lower bound of a, possibly exponential-size, set. This addresses Theorem~\ref{Cohomology of curves}.

More generally for smooth, projective varieties $X$, the theory of \'etale cohomology, in particular the Kummer sequence, allows us to relate the group $\mathrm{H}^{1}(X, \mathbb{Z}/\ell \mathbb{Z})\simeq \mathrm{Pic}^{0}(X)[\ell]^{\vee}$ to the $\ell$-torsion in the Picard scheme of $X$. Define the $\ell$-adic versions of the cohomology,
$$
\mathrm{H}^{1}(X, \mathbb{Z}_{\ell}):=\lim_{\leftarrow j} \mathrm{H}^{1}(X, \mathbb{Z}/\ell^{j}\mathbb{Z}) \ \ \text{and} \ \ \mathrm{H}^{1}(X, \mathbb{Q}_{\ell}):=\mathrm{H}^{1}(X, \mathbb{Z}_{\ell})\otimes \mathbb{Q}_{\ell} \,.
$$
By an application of the weak-Lefschetz theorem (Theorem~\ref{weaklefs}), we notice that to compute $P_{1}(X/\mathbb{F}_{q}, T)$, it is sufficient to compute $P_{1}(Y/\mathbb{F}_{q}, T)$ where $Y$ is a smooth projective {\em surface} obtained by successively taking smooth hyperplane sections of $X$. 
By Algorithm~\ref{alg:lef}, we produce a Lefschetz pencil of hyperplane sections on $Y$, denoted $(H_{t})_{t\in \mathbb{P}^{1}}$, with $Y_{t}:=H_{t}\cap Y$ being smooth {\em curves}, for $t$ in an open dense subscheme $U_{0}\subseteq \mathbb{P}^{1}$. Denote $U:=U_{0}\times \overline{\mathbb{F}}_{q}$.

This procedure gives us (implicitly) a morphism\footnote{$\tilde{Y}$ is a smooth projective surface obtained by blowing up $Y$ along $\Delta\cap Y$, where $\Delta$ is the axis of the pencil.} $\pi:\tilde{Y}\rightarrow \mathbb{P}^{1}$, whose fibre at any $t\in \mathbb{P}^{1}$ is $Y_{t}$. By the Leray spectral sequence, we have $$\mathrm{H}^{1}(Y, \mathbb{Q}_{\ell})\simeq \mathrm{H}^{1}(\tilde{Y}, \mathbb{Q}_{\ell})\simeq \mathrm{H}^{0}(\mathbb{P}^{1}, \mathcal{F}),$$ where $\mathcal{F}:=R^{1}\pi_{\star}\mathbb{Q}_{\ell}$ is the \'etale sheaf, on the projective line, obtained by the first direct image relative to $\pi$. Further, by the proper base-change theorem, we have for any $t\in \mathbb{P}^{1}$, the stalk $\mathcal{F}_{t}\simeq \mathrm{H}^{1}(Y_{t}, \mathbb{Q}_{\ell})$. We notice that $\mathcal{F}\vert_{U}$ is a locally constant sheaf on $U$ and has as a subsheaf $\mathcal{E}\subset \mathcal{F}\vert_{U}$, the sheaf of vanishing cycles. The sheaf $\mathcal{E}$ is locally constant and of rank (say) $2r$.

We prove an effective version (Theorem~\ref{effgcd}) of Deligne's `th\'eor\`eme du pgcd' (``polynomial gcd theorem'' from the celebrated work \cite{Weilii}). In particular, we show that there exists an extension $\mathbb{F}_{Q}/\mathbb{F}_{q}$ of bounded degree such that we can recover (with high probability) $P_{1}(Y/\mathbb{F}_{Q}, T)$, merely from the curve-case polynomials $P_{1}(Y_{u_i}/\mathbb{F}_{Q}, T)$ with $u_{i}\in U(\mathbb{F}_{Q})$ chosen randomly, for $1\leq i\leq 2$; by computing their gcd. The Theorems~\ref{Theorem 1}, \ref{Theorem 3} and \ref{Theorem 2} follow from this. The ingredients are as follows. The hard-Lefschetz theorem (Theorem~\ref{hardlef}) states $$\mathrm{H}^{1}(Y_{u}, \mathbb{Q}_{\ell})=\mathrm{H}^{1}(Y, \mathbb{Q}_{\ell}) \oplus \mathcal{E}_{u}$$ for $u\in U$. We proceed to understand the action of the Frobenius at $u$ on $\mathcal{E}_{u}$, which for our purposes behaves as a `random group' contribution. The sheaf $\mathcal{E}_{\mathbb{Z}_{\ell}}\subset R^{1}\pi_{\star}\mathbb{Z}_{\ell}\vert_{U}$ of $\ell$-adic integral vanishing cycles on $U$ corresponds to a representation of the \'etale fundamental group $\rho: \pi_{1}(U_{0}, u) \rightarrow \mathrm{GL}(2r, \mathbb{Z}_{\ell})$ via its action on the stalk of $\mathcal{E}_{\mathbb{Z}_{\ell}}$ at $u$. We next study the geometric mod-$\ell$ monodromy $\overline{\rho}_{\ell}:\pi_{1}(U, u)\rightarrow \mathrm{GL}(2r, \mathbb{F}_{\ell})$. Methods of Hall \cite{hall} imply that $\mathrm{im}(\overline{\rho}_{\ell})=\mathrm{Sp}(2r, \mathbb{F}_{\ell})$, the symplectic group, when $\ell$ is such that the $\mathrm{H}^{i}(Y, \mathbb{Z}_{\ell})$ are all torsion-free. An equidistribution theorem due to Katz dictates the proportion of Frobenius elements $F_{Q, v}\in \pi_{1}(U_{0}, u)$ for $v\in U(\mathbb{F}_{Q})$, whose image lies in a conjugacy-stable subset of the mod-$\ell$ arithmetic monodromy group. The error term therein, and an analysis of the proportion of matrices in the group of symplectic similitudes $\mathrm{GSp}(2r, \mathbb{F}_{\ell})$ with characteristic polynomial coprime to a given one; provide the reasonable bounds for $\ell$ and $Q$ to obtain our computational complexity result. The underlying torsion-bounds employ the work of Kweon \cite{kweon}, along with our good-reduction assumption for $\mathcal{X}_{0}$ at the prime $\mathfrak{p}$.

%

\section{Zeta function of curves}
\label{sec2}

In this section, we present an $\mathrm{AM}\cap\mathrm{coAM}$ protocol for certifying the zeta function of a 
curve $C/\mathbb{F}_q$. We assume the input to be a smooth, projective, absolutely irreducible curve $C_{0}\subset \mathbb{P}^{N}$ of genus $g>0$ and degree $\delta$, presented as a system of homogeneous polynomials $f_1,\ldots,f_m$ with coefficients in $\mathbb{F}_{q}$ and of degree $\leq d$. Denote by $C$ the base change to the algebraic closure $\overline{\mathbb{F}}_{q}$. We begin with the preliminary subsections~\ref{Preliminaries} and ~\ref{subsec:arith} consisting of standard material. The $\mathrm{AM}\cap \mathrm{coAM}$ protocol of Theorem~\ref{Cohomology of curves} and its proof is presented in ~\ref{subsec:am}.
\subsection{Preliminaries}
\label{Preliminaries}
A \textit{divisor} $D$ on $C$ is a formal sum 
$
D \;=\; \sum_{i=1}^{r} n_{i}P_{i}
\,,$
where $P_i\in C(\overline{\mathbb{F}}_{q})$ and $n_{i}\in \mathbb{Z}\setminus\{0\}$. The set of points $P_{i}$ occurring in the sum above is called the \textit{support} of $D$. The sum $\sum_i n_i$ is called the \textit{degree} of $D$.

We denote the group of divisors by $\mathrm{Div}(C)$ and the subgroup of degree zero divisors by $\mathrm{Div}^0(C)$. Let $K$ denote the function field of $C$. We have a map $\mathrm{div}: K^{*}\hookrightarrow \mathrm{Div}^{0}(C)$, sending a function to the sum of its zeros and poles. The image of this map is called the subgroup of \textit{principal divisors}, denoted $\mathrm{Div}^{\mathrm{pr}}(C)$. We call a divisor $D$ \textit{effective}, if $n_i\geq 0$ for all $i$, which we denote by $D\geq 0$.
 
\begin{definition}
    There exists an abelian variety (of dimension $g$) called the \textit{Jacobian}, denoted $\mathrm{Jac}(C)$, whose $\overline{\mathbb{F}}_{q}$-rational points correspond to elements of the quotient group $\mathrm{Div}^{0}(C)/\mathrm{Div}^{\mathrm{pr}}(C)$. 
\end{definition} 
Let $D$ be a divisor on $C$. We recall the \textit{Riemann-Roch space} of $D$.
$$
\mathcal{L}(D) \,:=\,\{f\in K^{*} \ \vert \ \mathrm{div}(f)+D\geq 0\}\cup\{0\} \,.
$$
Further, denoting by $K_{C}$ the canonical divisor of $C$, the Riemann-Roch theorem states
    $$\dim \mathcal{L}(D) \,=\, \mathrm{deg}(D)+1-g+\dim \mathcal{L}(K_C-D) \,.$$
    
 Addition on the Jacobian is performed by using an effective Riemann-Roch theorem. However, in order to invoke algorithms \cite{HwaIer,abelardRiemannRoch} computing the Riemann-Roch spaces, we first reduce our curve to a {\em planar} model.




 In particular, we seek to find a curve $C'\subset \mathbb{P}^{2}$ birational to $C$, given by a homogeneous form $F$. A singular point $P\in C'$ is said to be a \textit{node} if it is an ordinary double point, i.e., has multiplicity two, with distinct tangents. A curve is \textit{nodal} if all its singularities are nodes. We recall \cite[IV.3.11]{hartshorne}.

\begin{lemma}[Planar model]
\label{planemodel}
    Let $C\subset \mathbb{P}^{N}$ be as above. There is a randomised algorithm that computes a nodal curve $C'\subset \mathbb{P}^{2}$ and a birational morphism $\phi:C\rightarrow C'$ that runs in time polynomial in $g\log q$.
\end{lemma}
\begin{proof}
We describe how to obtain an equation defining $C'$ algorithmically. The key idea is to choose a random point $O\in \mathbb{P}^{N}$, with $O\not\in C$, and project $C$ onto a hyperplane from $O$. For generic $O$ (lying outside any secant or tangent of $C$) and $N\geq 4$, the resulting map is an embedding. Repeating the process, we get a sequence of morphisms $C\rightarrow \mathbb{P}^{N-1}\rightarrow\cdots \rightarrow \mathbb{P}^{3}$. The locus of `bad' projections forms a subvariety of $\mathbb{P}^{3}$ of dimension at most $2$, with degree bounded by a polynomial in $\delta:= \deg(C)$. Hence, this locus can be avoided with high probability at the cost of a field extension of degree at worst poly($\delta$). Therefore, generically, by \cite[Theorem V.3.10]{hartshorne} for $O\in \mathbb{P}^{3}$, the image of the projection of $C$ from $O$ onto $\mathbb{P}^{2}$ has at worst nodal singularities. Denote by $\phi:C\rightarrow \mathbb{P}^{2}$ the composite morphism of all projections. It is a birational morphism with $\mathrm{deg}(\phi(C))\leq \delta$. Therefore, the polynomial $F$ cutting out $C'$ in $\mathbb{P}^{2}$ has total degree at most $\delta$. Writing the linear projection $\phi$ explicitly and computing the image of $\Theta(\delta^{2})$ many points $P_{i}\in C$, we can recover $F$ by a bivariate interpolation algorithm. Points on the curve can be sampled by the procedure below. \\ 

Sampling points in $C(\mathbb{F}_{q})$ (which exist after an extension) can be achieved in randomised polynomial time as follows. Consider an affine piece of $C$ in $\mathbb{A}^{N}$ (with coordinates $(y_{1}, \ldots, y_{N})$) by taking the complement of a hyperplane. Fixing a value of $y_{1}$ amounts to intersecting with a hyperplane in $\mathbb{A}^{N}$, giving a finite set of points. The Weil bound (see Theorem~\ref{weil} below) for $C$ guarantees that with high probability, after $2g\leq 4\delta^{2}$ fixings of $y_{1}$ in $\mathbb{F}_{q}$, the resulting zero-dimensional system has $\mathbb{F}_{q}$-rational points. Extracting them can be done in randomised polynomial-time by using the main result of \cite{lazard} for the $\mathbb{F}_{q}$ -root-finding of a zero-dimensional $N$-variate system. 
\end{proof}
We conclude this subsection with a statement of the Weil-Riemann hypothesis for curves \cite{WeilA,WeilB}.
\begin{theorem}[Weil]
\label{weil}
    $\vert\# C(\mathbb{F}_{q})-(q+1) \vert \leq 2g\sqrt{q}$ .
\end{theorem}

\subsection{Jacobian arithmetic}
\label{subsec:arith}
Recall the standard results showing that elements of $\mathrm{Jac}(C)(\mathbb{F}_q)$ can be presented concisely and divisor arithmetic therein can be performed efficiently. 
 We know by \cite[\S8]{serrAlgGrp} that $C$ injects into its Jacobian, by the choice of a rational point, which we call $\infty$.

\begin{lemma}[Reduced form] 
\label{reduced divisor representation}
  Given $D\in \mathrm{Jac}(C)$, $\exists \ 0\leq i\leq g$ and a unique {\em effective} divisor $E$ of degree $g-i$ such that $D = E-(g-i)(\infty)$ in $\mathrm{Jac}(C)(\overline{\mathbb{F}}_{q})$. 
\end{lemma}

\begin{proof}
By the Riemann-Roch theorem, we have $\dim \mathcal{L}(D+g(\infty))=1+\dim  \mathcal{L}(K_{C}-D-g(\infty))> 0.$ Iteratively, subtracting $\infty$ from the divisor $D+g(\infty)$, we choose the largest $0\leq i\leq g$ so that $\dim \mathcal{L}(D+(g-i)(\infty))$ is still positive. In particular, for such an $i$, we have $\dim \mathcal{L}(D+ (g-i)(\infty))=1.$ Thus, one gets a `unique' (upto a constant) rational function $f$ in the basis of $\mathcal{L}(D+(g-i\infty))$. Therefore, one obtains a unique effective divisor $E:= \mathrm{div}(f)+D+ (g-i)(\infty) \ge 0 \;,$ which is the same as saying $D= E-(g-i)(\infty)$ in the arithmetic of $\mathrm{Jac}(C)(\overline{\mathbb{F}}_{q})$. 
\end{proof}


We recall next a method to compute bases of Riemann-Roch spaces.
\begin{proposition}[Riemann-Roch basis]
\label{rrspace}
    Let $D$ be a divisor on a curve $C$ of degree and support-size $\le \delta$. A basis of the Riemann-Roch space $\mathcal{L}(D)$ can be computed efficiently in $O(\delta^{12}\log q)$ time.
\end{proposition}

\begin{proof}
After computing a plane model $\phi:C\rightarrow C'\subset \mathbb{P}^{2}$ one uses \cite[\S 2]{HwaIer} to compute the Riemann-Roch space of a divisor on the normalisation of $C'$ (which is isomorphic to $C$). While \cite{HwaIer} requires the singular points of $C'$ to lie over the base field (essentially to ensure an efficient resolution of singularities), this can be bypassed by using \cite{Kozen} instead. The complexity follows from \cite[$\S 2.5$]{HwaIer}. This strategy was also utilised in the algorithm of \cite[$\S 6$]{kedlaya2006quantum} as a preprocessing step to do basic arithmetic in the class group (=$\mathrm{Jac}(C)$). 
\end{proof}

Using Proposition~\ref{rrspace}, we can now check when a divisor of degree zero is trivial in the Jacobian. Recall that for $D\in \mathrm{Div}^{0}(C)$, we have $\dim \mathcal{L}(D)=1$ if and only if $D\in \mathrm{Div}^{\mathrm{pr}}(C)$. This implies the following.
\begin{lemma}[Zero test]
\label{princ}
    Given $D\in \mathrm{Div}^{0}(C)$, whether $D\in \mathrm{Div}^{\mathrm{pr}}(C)$ is testable in polynomial time. In other words zero-tests in $\mathrm{Jac}(C)$ can be performed in polynomial time.
\end{lemma}


Combining Lemma~\ref{reduced divisor representation}, Proposition~\ref{rrspace} and Lemma~\ref{princ}, one obtains a polynomial time algorithm to put a given divisor $D\in \mathrm{Jac}(C)(\mathbb{F}_{q})$ into reduced form. Indeed by Lemma~\ref{reduced divisor representation}, one knows that the support of $D$ can be chosen to be of size at most $\mathrm{poly}(g)$. Then, Proposition~\ref{rrspace} can be applied to obtain the effective divisor $E$ and the integer $i$, so that $D=E-(g-i)\infty$ is in reduced form as an element of $\mathrm{Jac}(C)$. 

\begin{remark}
    The points occurring in the support of the effective divisor $E$ associated to the reduced form of $D$ in the above description each lie in a $\mathrm{poly}(g)$ extension of $\mathbb{F}_{q}$. However, one never needs to go to an extension of $\mathbb{F}_{q}$ containing \textit{all} of them simultaneously (which may be exponentially large in degree). The issue is handled exactly the same way in \cite[\S 6]{kedlaya2006quantum}. See also \cite[\S 3]{HwaIer} for more on this \textit{implicit representation} of divisors used in their algorithm to do Jacobian arithmetic.
\end{remark}

\vspace{-1mm}
We are now ready to describe a randomised polynomial-time Algorithm \ref{addtwopoints} to compute the sum of two elements in $\mathrm{Jac}(C)$ in the canonical representation described above.
\vspace{-1mm}
\begin{algorithm}
\caption{Adding two points on the Jacobian}
\label{addtwopoints}

\begin{itemize}
    \item \textbf{Input:} Two divisors  $D_1=E_1-m_1(\infty)$ and $D_2=E_2-m_2(\infty)$ of degree zero, with $m_{1}, m_{2}\leq g$ lying in the Jacobian of a smooth projective curve $C/\mathbb{F}_{q}$, presented in the reduced form as per Lemma~\ref{reduced divisor representation}. 
    \item \textbf{Output:} $D_3=D_1+D_2$ as $D_3=E_3-m_3(\infty)$ where $E_3$ is effective of degree $m_3$.
\end{itemize}
\begin{algorithmic}[1]



\STATE {\em (Reduction loop)} For each $i$, compute $\mathcal{L}(D_1+D_2+(g-i)(\infty))$ using Proposition~\ref{rrspace}, starting from $i=0$. If  $\dim \mathcal{L}(D_1+D_2+(g-i)(\infty))=1$ then we get a unique effective divisor $E:= \mathrm{div}(f)+D_1+D_2+ (g-i)(\infty)$, where the representation of $\mathrm{div}(f)$ can be found in randomised polynomial-time \cite{lazard}. Choose the largest such $i$ and set $m_3=g-i$ and $E_3=E$.

\STATE  Output $E_{3}-m_3(\infty)$. 
\end{algorithmic}
\end{algorithm}

\subsection{$\mathrm{AM}$ protocol}
\label{subsec:am}
In this subsection, we present an $\mathrm{AM}\cap\mathrm{coAM}$ protocol to certify the order (and group structure) of $\mathrm{Jac}(C)(\mathbb{F}_{q})$. We then show how to certify the zeta function of $C$ using this. We first recall a result of Weil \cite[pp.70-71]{WeilA} which generalises a theorem of Hasse \cite[p.206]{Hasse} from elliptic curves ($g=1$) to abelian varieties ($g\ge1$).

\begin{proposition}[Hasse-Weil interval]
\label{Hasse-Weil} 
    For an abelian variety $A$ of dimension $g$ over the finite field $\mathbb{F}_{q}$, the number of $\mathbb{F}_q$-rational points is in the following range:
    $$
    (\sqrt{q}-1)^{2g} \;\leq \#A(\mathbb{F}_q) \; \leq (\sqrt{q}+1)^{2g} \,.
    $$
\end{proposition}
\noindent {\bf Reduced gap.} Given an input curve of genus $g$ we want to choose $q$ so that the above gap is small enough, namely, $( (\sqrt{q}+1)/ (\sqrt{q}-1) )^{2g} < 2$. In particular, we require
$$
1+\frac{2}{\sqrt{q}-1} \,<\, 2^{1/2g} \,=\, \mathrm{exp}\left(\frac{\log 2}{2g}\right)=1+\frac{\log 2}{2g}+\frac{(\log 2)^{2}}{8g^{2}}+\ldots
$$
Truncating, we notice that $q>(8g+1)^{2}$ suffices. 


\noindent {\bf Hash functions} are pseudorandom maps from large strings to small strings, in a way that minimizes {\em collision} as much as possible. 
Let $h:\{0,1\}^n\mapsto \{0,1\}^k;\,k\ll n$ be from a hash family. We require that for $X\in \{0,1\}^n$ and a random $Y\in \{0,1\}^k$, $\mathrm{Pr}_{h,Y}[h(X)=Y]=1/2^k$. One can show that, for a random $k\times n$ matrix $A$ over $\mathbb{F}_2$, and a random vector $b\in \{0,1\}^k$, $h:X\mapsto AX+b$ satisfies this property (see \cite[Theorem $8.15$]{AB}). Using this concept, Algorithm \ref{algzetacurve} is the $\mathrm{AM}\cap \mathrm{coAM}$ protocol to verify the Jacobian size, over $\mathbb{F}_Q \supseteq \mathbb{F}_q$ assuming $Q>(8g+1)^{2}$.
\vspace{-0.5mm}
\begin{algorithm}

\caption{Verifying the size and structure of the Jacobian of $C/\mathbb{F}_Q$}
\label{algzetacurve}

\textbf{Input:} A smooth projective {\em curve} $C\subset \mathbb{P}^{N}$ of genus $g$ and degree $\delta$, given by polynomials $(f_{i})_{1\leq i \leq m}$. A {\em candidate} integer $\mathcal{N}$ lying in the Hasse-Weil interval. Set $L$: $2^{L-1}< \mathcal{N} \leq 2^L$.

\begin{algorithmic}[1]
\STATE \textbf{Arthur}: Choose a random hash function $h: \{0,1\}^{2g\log Q}\rightarrow \{0,1\}^{L+1}$ by picking a matrix $A$ and a vector $b$ randomly as stated above. Pick a random $y\in \{0,1\}^{L+1}$ and send $(h,y)$ to Merlin as a {\em challenge}. {\bf Note:} Arthur could send $O(L)$ many such independently chosen pairs $(h,y)$ to reduce the error probability exponentially. Below, we use only one pair for the simplicity of exposition.

\STATE \textbf{Merlin}: \begin{itemize}
    \item Pick $r$ generators $D_i \in \mathrm{Jac}(C)(\mathbb{F}_{Q})$ ($i\in [r]$) such that $$\mathrm{Jac}(C)(\mathbb{F}_{Q}) \;\simeq\; \langle D_{1}\rangle \times \ldots \times \langle D_{r}\rangle$$ with each $D_{i}$ of order $n_{i}$, with $n_{i}\vert n_{i+1}$ and $\prod_{i=1}^{r}n_{i}=\mathcal{N}$. Each $D_{i}$ is presented in canonical form as $D_i=E_i-m_{i}(\infty)$, with $E_i$ effective of degree $m_{i}$. The divisors $E_{i}$ in turn are presented as a sum of $\overline{\mathbb{F}}_{Q}$ -- rational points of $C$, each defined over an extension of $\mathbb{F}_{Q}$ of degree at most $\mathrm{poly}(g)$ thanks to Lemma~\ref{reduced divisor representation}. 
    

    \item Send a {\em response} consisting of $r$ quadruples $\{(c_i,D_i,n_i,P_i)\}_{1\leq i\leq r}$ with the claim that the divisor $\sum_{i}c_{i}D_{i}=:x$, for $c_{i}\in \mathbb{Z}/n_{i}\mathbb{Z}$, satisfies $h(x)=y$. Every $P_i$ is a set of pairs: each consisting of a prime factor of $n_{i}$ and the corresponding exponent in its factorisation.
    \end{itemize} 

\STATE \textbf{Arthur}:\begin{itemize}
    \item Check whether $D_i$ indeed represents a point in $\mathrm{Jac}(C)(\mathbb{F}_{Q})$. This is done by evaluating the Frobenius $F_{Q}$ on $D_i=E_i-m_{i}(\infty)$ and checking for invariance. If not, $\textit{Reject}$. 
    \item Check the factorization data $P_i$ of each $n_i$. Check the order $n_i$ as follows: verify $n_iD_i=0$, and for each distinct prime factor $p_{i, j}$ of $n_i$, verify $(n_i/p_{i,j})D_i\ne 0$. Check that $\mathcal{N}=\prod_{i=1}^r n_i$. If a check fails, $\textit{Reject}$. Calculate $x=\sum_{i}c_{i}D_{i}$.
    \item Check $h(x) = h\left(\sum_i c_iD_i\right) = y$, if yes then $\textit{Accept}$; otherwise $\textit{Reject}$. All the checks can be easily performed by Arthur using: basic arithmetic, or Algorithm \ref{addtwopoints}, combined with the standard trick of repeated-doubling.
\end{itemize}
\end{algorithmic}
\end{algorithm}
\vspace{-2mm}
\begin{lemma}[Probability of Algorithm \ref{algzetacurve}]
\label{Probability bound}
    In Algorithm \ref{algzetacurve} (given candidate $\mathcal{N}$), if $\#\mathrm{Jac}(C)(\mathbb{F}_Q)=\mathcal{N}$, then Arthur accepts with probability $>2/3$. Else, Arthur {\em rejects} with probability $>2/3$. 
\end{lemma}
\begin{proof} We adapt the protocol from \cite[\S 9.4]{AB}.
Let $S\subset \{0,1\}^{2g\log Q}$ denote the set $\mathrm{Jac}(C)(\mathbb{F}_{Q})$ with the elements written as binary strings. Let $\mathcal{G}$ be the group generated by the divisors $D_i$'s that Merlin provided. Suppose it has size $\mathcal{N}$, as Merlin claimed. In particular, $\mathcal{G}=S$ as we have made the Hasse-Weil `gap' small enough so that only a unique multiple of $\mathcal{N}$ can lie in that interval. For a random $y\in \{0,1\}^{L+1}$ and a random hash function $h$ (chosen from a uniform distribution over matrices $A$ and vectors $b$ such that $h:x\mapsto Ax+b$), the probability that there is an $x\in \mathcal{G}=S$, such that $h(x)=y$ is 
\begin{align}
 \mathrm{Pr} [\exists x\in \mathcal{G}=S, \ h(x)=y] &\geq\;  \binom{\# S}{1}\cdot\frac{1}{2^{L+1}} - \binom{\# S}{2}\cdot\frac{1}{2^{2(L+1)}} \;>\; \frac{\#S}{2^{L+1}}-\frac{(\#S)^{2}}{2^{2(L+1)+1}}  \nonumber\\ 
 &>\; \frac{\# S}{2^{L+1}}\left(1- \frac{\#S}{2^{L+2}}\right) \;\ge\; 0.75\cdot\frac{\# S}{2^{L+1}} \;.\label{prob-high}
\end{align}
from the inclusion-exclusion-principle, and applying the inequality $2^{L-1}< \# S=\mathcal{N} \leq 2^{L}$. 

Conversely, suppose $\# S \ne \mathcal{N}$, as Merlin bluffed (so, $\mathcal{G}\ne S$). Since Arthur checked that the product of the orders of the divisors $D_i$'s equals $\mathcal{N}$, we deduce that $\#\mathcal{G}\le \#S/2$ (as the order of the subgroup $\mathcal{G}$ properly divides that of the group $S$). 
So, simply by the union-bound we get
\begin{align}
\mathrm{Pr} [\exists x\in \mathcal{G}, \ h(x)=y] &\le\;  \binom{\# \mathcal{G}}{1}\cdot\frac{1}{2^{L+1}}  
\;\le\; 0.5\cdot\frac{\# S}{2^{L+1}} \;.\label{prob-low}
\end{align}
Thus, Eqns.\ref{prob-high}-\ref{prob-low} have a noticeable difference in the probability estimate.
Now, we can repeat, with Arthur choosing several $(h,y)$ pairs, take the `majority vote', and use the Chernoff bound \cite[\S 7.4.1]{AB}. This {\em amplification} trick brings the probabilities above $2/3$ (in Eqn.\ref{prob-high}) and below $1/3$ (in Eqn.\ref{prob-low}) respectively. The number of repetitions will be inverse-polynomial in $\#S/2^{L+1} > 1/4$; which is only a constant blowup in our time complexity.
\end{proof}
\begin{remark}
    The steps of Merlin require exponential resources (i.e.~Step 2), so we do not know how to compute them in polynomial-time in practice. The purpose is to only provide a concise certificate, using which Arthur can verify the Jacobian-size efficiently and reliably (with high probability).
\end{remark}

\begin{lemma}[Complexity of Algorithm \ref{algzetacurve}]\label{Time}
    Arthur's verification algorithm runs in randomised polynomial-time. 
\end{lemma}
\begin{proof}
    Step $1$ simply involves addition and multiplication of binary matrices of size $\text{poly}(g\log Q)$, so can be accomplished in $\text{poly}(g\log q)$ time.     
    In Step $3$, since the number of prime factors of any integer $n$ is $O(\log n)$, the prime factor checking computation can be performed in $\text{poly}(\log\mathcal{N})$ time. Applying the Hasse-Weil bound, this is in fact $\text{poly}(g\log q)$ time.     
    For the Jacobian arithmetic, Arthur uses Algorithm \ref{addtwopoints} and repeated-doubling. 
    This sums up the complexity of our protocol to $\text{poly}(g, \log q)$-time.
\end{proof}

The zeta function is intimately connected to the order of the Jacobian. From \cite[\S 8]{kedlaya2006quantum}:
\begin{lemma}[Count to zeta function]
\label{ked1}
    Assume we are given $\#\mathrm{Jac}(C)(\mathbb{F}_{q^{j}})$, for every $1\leq j \leq \mathrm{max}(18, 2g)$. Then, $P_1(C/\mathbb{F}_{q}, T)$ can be reconstructed from these counts, in poly($g\log q$)-time.
\end{lemma}
Kedlaya \cite[\S 8]{kedlaya2006quantum} also shows the following, connecting the zeta function of a larger Frobenius  to that of a {\em smaller} Frobenius.
\begin{lemma}[Base zeta function]
\label{ked2}
    Let primes $m_{1}, m_{2}$ with $m_{1}<m_{2}$, be such that $m_{j}-1$ is divisible by a prime greater than $2g$, for $j\in\{1,2\}$. Assume further that $q^{m_{1}}>(8g+1)^{2}$. Then, $P_1(C/\mathbb{F}_{q}, T)$ can be recovered from $P_1(C/\mathbb{F}_{q^{m_j}}, T)$, $j\in\{1,2\}$, in time polynomial in $g\log q$. 
    
    Further, the existence of such $m_{1}, m_{2}$ bounded by a polynomial in $g\log q$ is guaranteed. \footnote{by \cite[Theorem 1.2]{harman}}
\end{lemma}

\begin{proof}[Proof of Theorem~\ref{Cohomology of curves}]
\label{pf1.4}
 Using Algorithm~\ref{algzetacurve}, we can verify the structure of $\mathrm{Jac}(C)(\mathbb{F}_{Q})$ for any $Q>(8g+1)^{2}$. This implies $P_1(C/\mathbb{F}_{q}, T)$ can be certified by first certifying $P_1(C/\mathbb{F}_{q^{m_{1}}}, T)$ and $P_1(C/\mathbb{F}_{q^{m_{2}}}, T)$  and next applying Lemma~\ref{ked2}. Each $P_1(C/\mathbb{F}_{q^{m_{j}}}, T)$ can be computed, uniquely, using the counts $\#\mathrm{Jac}(C)(\mathbb{F}_{q^{im_{j}}})$, for $1\leq i\leq \mathrm{max}(18, 2g)$, by Lemma~\ref{ked1}. This completes the proof of the first part of the theorem, verifying the zeta function.
 
        
\smallskip {\em Group structure.}    In the second part of the theorem statement, suppose a candidate $G$ has been provided via additive generators $\{A_1,\ldots, A_r\}$, with each $A_{i}$ of order $n_i$ such that $G$ decomposes as a direct sum of the subgroups $\langle A_i\rangle$, where $n_i\mid n_{i+1}$. We need to verify whether $\mathrm{Jac}(C)(\mathbb{F}_q)\simeq G$. For this, Merlin first convinces Arthur of the structure of $\mathrm{Jac}(C)(\mathbb{F}_Q)$, and provides the additive generators for $Q>(8g+1)^{2}$. Using this, Arthur can compute $P_{1}(C/ \mathbb{F}_{q}, T)$, thereby obtaining the count $\#\mathrm{Jac}(C)(\mathbb{F}_{q})=P_{1}(C/\mathbb{F}_{q}, 1)$. For the subgroup $\mathrm{Jac}(C)(\mathbb{F}_q)\subset \mathrm{Jac}(C)(\mathbb{F}_{Q})$, Merlin presents divisors $D_i$ with support in $C(\mathbb{F}_{Q})$, that are candidates corresponding to each $A_i$. Arthur first checks whether the $D_i$ all belong to $\mathrm{Jac}(C)(\mathbb{F}_q)$ (by evaluating the $q$-Frobenius on them and verifying invariance). Next, Arthur verifies the independence of the $D_{i}$ as in Algorithm~\ref{algzetacurve}. This provides a lower bound for $\# G$. Comparing it with the verified count $\#\mathrm{Jac}(C)(\mathbb{F}_{q})$ certifies the structure. The proof then follows from Lemmas \ref{Probability bound}-\ref{Time}.
\end{proof}

\section{Surfaces: Vanishing cycles, monodromy, and equidistribution}
\label{sec-technical}
This section is devoted to the technical background necessary to prove our main theorems in the higher dimensional case. In particular,~\ref{sec-VC} reviews the theory of vanishing cycles on a surface, including a statement of the hard-Lefschetz theorem in this case and the general `gcd theorem' of Deligne. Next, in ~\ref{sec-mono}, the Picard-Lefschetz formulas and the ($\ell$-adic and mod-$\ell$) monodromy  of a Lefschetz pencil of hyperplane sections on a surface are discussed including torsion bounds, followed by the statement of an equidistribution result of Katz. Finally in~\ref{sec-symp}, we briefly review symplectic groups over finite fields and deduce a probability estimate that we later use to prove  an effective version of Deligne's gcd theorem.

\subsection{Vanishing cycles}

\label{sec-VC} 
In this subsection, we give a brief overview of the theory of vanishing cycles associated to a surface fibred as a Lefschetz pencil over $\mathbb{P}^{1}$. Then, we discuss the `hard-Lefschetz theorem' and some implications for the first \'etale cohomology. Finally, we wrap with a statement of Deligne's `th\'eor\`eme du pgcd', which enables us to recover the characteristic polynomial of Frobenius, acting on the first cohomology, from its action on the cohomology of the fibres. 

Let $X_{0}$ be a smooth, projective, geometrically irreducible surface over the finite field $\mathbb{F}_{q}$ of characteristic $p>0$. Denote by $X$ the base change to the algebraic closure. Assume we have a Lefschetz fibration $\pi:\tilde{X}\rightarrow \mathbb{P}^{1}$ following Algorithm~\ref{alg:lef}. As usual, we let $Z\subset \mathbb{P}^{1}$ denote the set giving rise to singular fibres (nodal curves), and let $U$ denote its complement. Let $X_{\eta}$ be the generic fibre of $\pi$. It is a smooth curve of genus $g$ over the function field of $\mathbb{P}^{1}$. 

Let $\ell$ be an odd prime, coprime to $p$. Consider the sheaf $\mathcal{F}^{\ell}:=R^{1}\pi_{\star}\mu_{\ell}$ on $\mathbb{P}^{1}$. By the proper base-change theorem, we have that its stalk at a  point ${u}\rightarrow \mathbb{P}^{1}$ is the group $\mathrm{H}^{1}(X_{u}, \mu_{\ell})\simeq \mathrm{Pic}^{0}(X_{u})[\ell]$.
 Further, we know that $\mathcal{F}^{\ell}\vert_{U}$ is a locally constant sheaf of rank $2g$ on $U$. We seek to understand the behaviour of $\mathcal{F}^{\ell}$ at points $z\in Z$. Let $X'_{z}\rightarrow X_{z}$ be the normalisation (which has genus $g-1$) of such a singular fibre, and denote by $V_{z}$, the kernel of the map $\mathcal{F}^{\ell}_{z}\rightarrow \mathrm{Pic}^{0}(X'_{z})[\ell]$. We call $V_{z}$ the group of \textit{vanishing cycles} at $z$. We now recall a collection of results from \cite[V.3]{milne1980etale}.

\begin{proposition}
\label{mil}
With the above setup, the following are true:

\begin{itemize}        
        
        \item For any $u\in \mathbb{P}^{1}$ there exists a cospecialisation map $\mathcal{F}^{\ell}_{u}\rightarrow \mathcal{F}^{\ell}_{\eta}$ which is an isomorphism if and only if $u\in U$. 
        \item If $z\in Z$, the cospecialisation map $\mathcal{F}^{\ell}_{z}\rightarrow \mathcal{F}^{\ell}_{\eta}$ is an injection. In particular $\mathcal{F}^{\ell}_{z}\simeq (\mathbb{Z}/\ell \mathbb{Z})^{2g-1}$. Further, $V_{z}$ is the exact annihilator of $\mathcal{F}^{\ell}_{z}$ under the Weil-pairing map
        $$
        \langle \cdot,\cdot\rangle: \mathcal{F}^{\ell}_{\eta}\times \mathcal{F}^{\ell}_{\eta}\longrightarrow \mu_{\ell}(\overline{\mathbb{F}}_{q}) \,.
        $$
        
        \item $\mathcal{F}^{\ell}$ is tamely ramified at all $z\in Z$. 
    \end{itemize}
\end{proposition}

In particular, for $z\in Z$, we have  $V_{z}\simeq \mathbb{Z}/\ell \mathbb{Z}$ \footnote{we omit Tate-twists by fixing an isomorphism $\mathbb{Z}/\ell \mathbb{Z}\simeq \mu_{\ell}(\overline{\mathbb{F}}_{q})$ and choosing a generator for the group of roots of unity.}, and we denote by $\delta_{z}$, the element that maps to $1$. We may also identify $\delta_{z}$ with its image under the map $\mathcal{F}^{\ell}_{z}\rightarrow \mathcal{F}^{\ell}_{\eta}$, and call $\mathcal{E}^{\ell}(X_{\eta})$ the subspace generated by all the $\delta_{z_i}$ in $\mathcal{F}^{\ell}_{\eta}$ for $z_i\in Z$. By the cospecialisation map, we refer to the corresponding subspace generated in $\mathcal{F}^{\ell}_{u}$ for $u\in U$ by $\mathcal{E}^{\ell}(X_{u})$. By passage to the limit and tensoring, we also obtain the $\mathbb{Q}_{\ell}$-vector space of vanishing cycles $\mathcal{E}(X_{u})$. Moreover, there exists a locally constant subsheaf $\mathcal{E}\subset R^{1}\pi_{\star}\mathbb{Q}_{\ell} \vert_{U}
$, called the \textit{sheaf of vanishing cycles} with stalk $\mathcal{E}_{u}=\mathcal{E}(X_{u})$ for $u\in U$. We now recall the `hard-Lefschetz' theorem for surfaces, which measures precisely the discrepancy between $\mathrm{H}^{1}(X_{u}, \mathbb{Q}_{\ell})$ and $\mathrm{H}^{1}(X, \mathbb{Q}_{\ell})$.

\begin{theorem}[Hard-Lefschetz]
\label{hardlef}
We have the decomposition 
$$
\mathrm{H}^{1}(X_{u}, \mathbb{Q}_{\ell})\simeq \mathrm{H}^{1}(X, \mathbb{Q}_{\ell})\oplus \mathcal{E}_{u}
$$
with respect to the symplectic pairing. In particular, $\mathrm{H}^{1}(X, \mathbb{Q}_{\ell})\simeq \mathcal{E}_{u}^{\perp}$ when viewed as a subspace of $\mathrm{H}^{1}(X_{u}, \mathbb{Q}_{\ell})$ under the weak-Lefschetz map.
\end{theorem}
The general result is a deep theorem of Deligne \cite[4.3.9]{Weilii}. The surface case is easier to handle and is done in \cite[2.A.10]{kleiman}. We conclude this subsection with a statement of Deligne's `th\'eor\`eme du pgcd' \cite[4.5.1]{Weilii}.

Let $X_{0}/\mathbb{F}_{q}$ now be a smooth, projective, geometrically irreducible variety of dimension $n$ and let $X$ be the base change to the algebraic closure.

\begin{theorem}[Le th\'eor\`eme du pgcd]
\label{gcd}
    Let $(X_{t})_{t\in \mathbb{P}^{1}}$ be a Lefschetz pencil of hypersurface sections of degree $d\geq 2$ on $X$. Then $P_{n-1}(X/\mathbb{F}_{q}, T)$ is the least common multiple of all polynomials $f(T)\in \mathbb{C}[T]$, satisfying the condition that for any $t\in \mathbb{F}_{q^{r}}$ such that $X_{t}$ is smooth, the polynomial \footnote{if $f(T)=\prod_{j}(1-\alpha_{j}T)$, then $f(T)^{(r)}:=\prod_{j}(1-\alpha_{j}^{r}T)$.} $f(T)^{(r)}$ divides $P_{n-1}(X_{t}/\mathbb{F}_{q^{r}}, T)$.
    
\end{theorem}
Deligne derived the above as a consequence of his proof of the Weil-Riemann hypothesis and hard-Lefschetz theorem for $\ell$-adic cohomology. Theorem~\ref{gcd} has been used by Katz-Messing in \cite{katzmessing} to deduce the same facts for any Weil cohomology theory. The theorem was also used by Gabber in \cite{gabber} to show torsion-freeness of the integral $\ell$-adic cohomology for smooth projective varieties for `almost all' $\ell$.

\subsection{Monodromy and equidistribution}
\label{sec-mono}
In this subsection, we introduce the notion of monodromy in a Lefschetz pencil. We then recall a big mod-$\ell$ monodromy result, obtained by an adaptation of work of Hall. Finally, we state a version of Deligne's equidistribution theorem \cite[3.5.3]{Weilii} due to Katz.
As before, let $\pi:\tilde{X}\rightarrow \mathbb{P}^{1}$ be a Lefschetz pencil of curves on a smooth, projective surface $X$. We denote by $U_{0}\subset \mathbb{P}^{1}$ the locus parameterising smooth fibres (of genus $g$) and by $U=U_{0}\times \overline{\mathbb{F}}_{q}$. Let $Z=\mathbb{P}^{1}\setminus U$ be the finite set parameterising the critical fibres. Write $\mathcal{F}=R^{1}\pi_{\star}\mathbb{Q}_{\ell}$ and $\mathcal{F}^{\ell}=R^{1}\pi_{\star}\mu_{\ell}$ for the respective direct-image sheaves.

Let $u\in U$ be a geometric point. The \textit{arithmetic \'etale fundamental group} (see \cite{murre} for the definition) $\pi_{1}(U_{0}, u)$ acts on $\mathcal{F}^{\ell}_{u}$ and by passage to the limit, on $\mathcal{F}_{u}$. This latter representation restricted to the \textit{geometric \'etale fundamental group} $\pi_{1}(U, u)$ is called the \textit{monodromy} of the pencil. Since $\mathcal{F}$ is tamely ramified, the action of $\pi_{1}(U, u)$ factors through the \textit{tame fundamental group}\footnote{essentially classifying finite \'etale covers of $U$ tamely ramified over $Z$.} $\pi_{1}^{t}(U, u)$. By a theorem of Grothendieck \cite[182-27]{fga}, $\pi_{1}^{t}(U, u)$ is generated topologically by $\#Z$ elements $\sigma_{i}$ for each $z_{i}\in Z$, satisfying the relation $\Pi_{i=1}^{\#Z}\sigma_{i}=1$. The Picard-Lefschetz formulas make this action explicit. See \cite[Ch V, Theorem 3.14]{milne1980etale} or \cite[III.4.3]{fk} for a proof.

\begin{proposition}[Picard-Lefschetz formulas]
For any $\gamma\in \mathcal{F}^{\ell}_{u}$, we have 
    \begin{equation}
    \label{eqn:piclef}
    \sigma_{i}(\gamma) \;=\; \gamma-\epsilon_{i}\cdot \langle\gamma, \delta_{z_i}\rangle\cdot \delta_{z_i} \;,
    \end{equation}
where for a uniformising parameter $\theta_{i}$ at $z_{i}$, we have $\sigma_{i}(\theta_{i}^{1/\ell})=\epsilon_{i}\cdot \theta_{i}^{1/\ell}.$ 
\end{proposition}
Clearly, the monodromy action respects the symplectic pairing. By the hard-Lefschetz theorem, we know that $\mathrm{H}^{1}(X_{u}, \mathbb{Q}_{\ell})\simeq \mathrm{H}^{1}(X, \mathbb{Q}_{\ell})\oplus \mathcal{E}_{u}$, with $\mathrm{H}^{1}(X, \mathbb{Q}_{\ell})=\mathcal{E}_{u}^{\perp}$. In particular, $\pi_{1}(U, u)$ acts trivially on $\mathrm{H}^{1}(X, \mathbb{Q}_{\ell})$, implying that the monodromy action factors through $\mathrm{Sp}(\mathcal{E}_{u})$, the group of symplectic transformations of the vector space $\mathcal{E}_{u}$. We know \cite[5.10]{Weili} that the image of $\pi_{1}(U, u)$ is open and Zariski-dense in $\mathrm{Sp}(\mathcal{E}_{u})$. Further, by the conjugacy of vanishing cycles, we also know $\pi_{1}(U, u)$ acts absolutely irreducibly on $\mathcal{E}_{u}$.

One seeks a version of the above to compute the mod-$\ell$ geometric monodromy for certain equidistribution estimates coming from Theorem~\ref{katz}. Consider the torsion-free sheaf $R^{1}\pi_{\star}\mathbb{Z}_{\ell}$ of rank $2g$ on $U$. It has as subsheaf, the sheaf of integral $\ell$-adic vanishing cycles $\mathcal{E}_{\mathbb{Z}_{\ell}}\subset R^{1}\pi_{\star}\mathbb{Z}_{\ell}$ of rank, say, $2r$. This in turn, corresponds to  representations $\rho:\pi_{1}(U_{0}, u)\rightarrow \mathrm{GL}(2r, \mathbb{Z}_{\ell})$ and $\overline{\rho}=\rho\vert \pi_{1}(U, u)$. Let $\mathcal{V}:= \mathcal{E}_{\mathbb{Z}_{\ell}}\otimes_{\mathbb{Z}_{\ell}} \mathbb{F}_{\ell}$ be the lisse $\mathbb{F}_{\ell}$-sheaf  giving rise to, respectively, the mod-$\ell$ representations $\rho_{\ell}$ and  $\overline{\rho}_{\ell}$. There are multiple ways to show big mod-$\ell$ monodromy for `almost all primes $\ell$' (all but finitely many), but \cite[\S 4-6]{hall} gives a method that works for \textit{every} prime $\ell\geq 5$ invertible on the characteristic. However, the generic rank of the local system $\mathcal{V}$ is a priori dependent on $\ell$, and guaranteed to be $2r$ only when the cohomology groups $\mathrm{H}^{i}(X, \mathbb{Z}_{\ell})$ are all torsion-free. The following result appears to be known to Hall and Katz (\cite[pg 5]{hall} and \cite{hal}); for completeness, we provide a brief proof below.

\begin{theorem}[Big mod-$\ell$ monodromy]
\label{bigm}
If $\mathrm{H}^{i}(X, \mathbb{Z}_{\ell})$ are all torsion-free, we have $$G:=\overline{\rho}_{\ell}\left(\pi_{1}(U, u)\right)=\mathrm{Sp}(2r, \mathbb{F}_{\ell}).$$
\end{theorem}

\begin{proof}
As the groups $\mathrm{H}^{i}(X, \mathbb{Z}_{\ell})$ are all torsion-free, we know (analogously to the situation in \cite[Theorem 9.2]{katzreport}) that the hard-Lefschetz theorem holds with $\mathbb{F}_{\ell}$ -- coefficients, i.e., for a smooth hyperplane section $X_{u}$ obtained as a fibre of our Lefschetz pencil, we have $$\mathrm{H}^{1}(X_{u}, \mathbb{F}_{\ell})\simeq \mathrm{H}^{1}(X, \mathbb{F}_{\ell})\oplus \mathcal{V}_{u}.$$ We now show that the representation $\mathcal{V}_{u}$ of $\pi_{1}(U, u)$ is irreducible. Indeed if $W\subset \mathcal{V}_{u}$ is a stable subspace, for any $\gamma\neq 0\in W$, we must have $\langle \gamma, \delta_{j}\rangle \neq 0$ for some $j$, as otherwise the Weil pairing would be degenerate on $\mathrm{H}^{1}(X_{u}, \mathbb{F}_{\ell})$. Therefore, by (\ref{eqn:piclef}), this implies $\sigma_{j}(\gamma)-\gamma=\epsilon_{j}\cdot \langle \gamma, \delta_{j}\rangle \cdot \delta_{j}\in W$. As the vanishing cycles are all conjugate under the action of $\pi_{1}(U, u)$ \cite[6.6]{katzlef}, this means $\delta_{i}\in W$ for all $i$, or $W=\mathcal{V}_{u}$.

We then invoke a theorem of Hall \cite[Theorem 3.1]{hall}, to conclude that the image is in fact the full symplectic group, due to the irreducibility and the transvections coming from the Picard-Lefschetz formulas.

\end{proof}

Let $X_{0}/\mathbb{F}_{q}$ now be obtained via good reduction from a smooth, projective, geometrically irreducible surface $\mathcal{X}_{0}$ over a number field $K$ at a prime $\mathfrak{p}$. We assume $\mathcal{X}_{0}\subset \mathbb{P}^{N}$ is of degree $D>0$ and given by the vanishing of homogeneous forms $f_{1}, \ldots, f_{m}$ each of degree $\leq d$. Denote $X:=X_{0}\times_{\mathbb{F}_{q}} \overline{\mathbb{F}}_{q}$ and $\mathcal{X}^{\mathrm{an}}:=\mathcal{X}_{0}\times_{K} \mathbb{C}$, equipped with the complex analytic topology. One has the following.
\begin{proposition}

\label{ellbound}
    There exists a prime $\ell$ with $(4D)^{4} \leq \ell\leq 2^{11}D^{N^{2}}$, coprime to $q$, such that $\mathrm{H}^{i}(X, \mathbb{Z}_{\ell})$ are all torsion-free for $0\leq i\leq 4$. 
\end{proposition}
 \begin{proof}
    Since $\mathcal{X}_{0}$ is a surface, we know, for Betti (co)homology $$\mathrm{H}_{1}(\mathcal{X}^{\mathrm{an}}, \mathbb{Z})_{\mathrm{tors}}\simeq (\pi_{1}(\mathcal{X}^{\mathrm{an}})^{\mathrm{ab}})_{\mathrm{tors}}\simeq \mathrm{H}^{2}(\mathcal{X}^{\mathrm{an}}, \mathbb{Z})_{\mathrm{tors}}\simeq \mathrm{H}^{3}(\mathcal{X}^{\mathrm{an}}, \mathbb{Z})_{\mathrm{tors}},$$ by Poincar\'e duality, the Hurewicz theorem and the universal coefficient theorem. Further, Kweon \cite[Corollary 5.4]{kweon} shows the following, as a consequence of bounds for torsion in the N\'eron-Severi group
    $$
    \prod_{\ell \neq p}\#\mathrm{H}^{2}(X, \mathbb{Z}_{\ell})_{\mathrm{tor}}\leq 2^{D^{N^{2}}+2N\log N}\leq 4^{D^{N^{2}}}.
    $$
The result follows as a consequence of analysing the growth of the primorial function \cite[XXII]{hw}
    $$
   2^{n/2} \leq \pi^{!}(n):=\prod_{\ell \ \text{prime}}^{ \ell \leq n}\ell \leq 4^{n}
    $$
    and applying standard comparison theorems for \'etale cohomology. 
\end{proof}

We close this subsection with the statement of a powerful Chebotarev-type equidistribution theorem due to Katz \cite[Theorem 9.7.13]{kasar}.

Let $U_{0}/\mathbb{F}_{q}$ be a smooth, affine, geometrically irreducible curve. Let $U$ be the base change to the algebraic closure. Pick a geometric point $u\rightarrow U$, lying over a closed point $u_{0}\in U(\mathbb{F}_{q})$ and denote by $\overline{\pi}_{1}:=\pi_{1}(U, u)$ the geometric \'etale fundamental group. Let $\pi_{1}$ denote the arithmetic fundamental group $\pi_{1}(U_{0}, u)$.

For any closed point $v\in U(\mathbb{F}_{q})$, there exists an element $F_{q, v}\in \pi_{1}$ well-defined upto conjugacy, called the \textit{Frobenius element} at $v$. It is defined as follows. Writing $v=\mathrm{Spec}(\mathbb{F}_{q})\rightarrow U$, we obtain an induced map of fundamental groups
$$
\mathrm{Gal}(\overline{\mathbb{F}}_{q}/ \mathbb{F}_{q})\rightarrow \pi_{1}(U_{0}, v) \simeq \pi_{1}.
$$
The element $F_{q, v}\in \pi_{1}$ is simply the image in $\pi_{1}$ of the frobenius element in $\mathrm{Gal}(\overline{\mathbb{F}}_{q}/ \mathbb{F}_{q})$ under the composition of the above morphisms.

Given a map $\rho: \pi_{1}\rightarrow G$ to a finite group, and a conjugacy-stable subset $C\subset G$, we seek to understand the proportion of points $v\in U(\mathbb{F}_{q^{w}})$ such that $\rho(F_{q^{w}, v})$ lies in $C$. The following is \cite[Theorem 4.1]{chav}.

\begin{theorem}[Katz]
\label{katz}
Assume there is a commutative diagram

\[
\begin{tikzcd}
1 \arrow{r}  & \overline{\pi}_{1} \arrow {r} \arrow{d}[swap]{\overline{\rho}} & \pi_{1} \arrow{r} \arrow{d}{\rho} & \hat{\mathbb{Z}} \arrow{r} \arrow{d}{1\mapsto -\gamma} & 1 \\
1 \arrow{r} & \overline{G} \arrow{r} & G \arrow{r}{\mu} & \Gamma \arrow{r} & 1
\end{tikzcd}
\]
where $G$ is a finite group, $\Gamma$ is abelian, $\overline{\rho}$ is surjective and tamely ramified. Let $C\subset G$ be stable under conjugation by elements of $G$. Then
$$
\left\vert \frac{\# \{v\in U(\mathbb{F}_{q^{w}}) \mid \rho(F_{q^{w}, v})\in C\}}{\#U(\mathbb{F}_{q^{w}})}-\frac{\#(C\cap G^{\gamma^{w}})}{\# \overline{G}}  \right\vert \,\leq\, \vert \chi(U)\vert \frac{\# G \sqrt{q^{w}}}{\#U(\mathbb{F}_{q^{w}})}\,,
$$
where $G^{\gamma^{w}}=\mu^{-1}(\gamma^{w})$ and $\chi(U)=\sum_{i=0}^{1}(-1)^{i}\dim \mathrm{H}^{i}(U, \mathbb{Q}_{\ell})$ is the $\ell$-adic Euler-Poincar\'e characteristic of $U$.
\end{theorem}

\subsection{Symplectic groups over finite fields}
\label{sec-symp}
The goal of this subsection is to obtain a probability estimate (Lemma~\ref{symplectic}) for use in Theorem~\ref{effgcd}. Let $V$ be a vector space of rank $2r$, for $r\in \mathbb{Z}_{>0}$, over the finite field $\mathbb{F}_{\ell}$ of characteristic $\ell>0$, equipped with a {\em symplectic} (i.e., alternating, nondegenerate, bilinear) pairing $\langle \cdot, \cdot\rangle$. 
\begin{definition}
The group of \textit{symplectic similitudes}, $\mathrm{GSp}(2r, \mathbb{F}_{\ell})$ is defined as
$$
 \mathrm{GSp}(2r, \mathbb{F}_{\ell}):=\{A\in \mathrm{GL}(2r, \mathbb{F}_{\ell}) \ \vert \ \exists \ \gamma \in \mathbb{F}_{\ell}^{*}  \ \text{such that} \ \langle Av, Aw\rangle =\gamma \langle v, w\rangle \ \forall v,w\in V\}.
$$
For $A\in \mathrm{GSp}(2r, \mathbb{F}_{\ell})$, the associated $\gamma\in \mathbb{F}_{\ell}^{*}$ is called the \textit{multiplicator} of $A$. We denote by $\mathrm{GSp}(2r, \mathbb{F}_{\ell})^{\gamma}$ the subset of matrices with multiplicator $\gamma$. The matrices with multiplicator $\gamma=1$ form a subgroup known as the \textit{symplectic group}, denoted $\mathrm{Sp}(2r, \mathbb{F}_{\ell})$.
We have the following exact sequence
$$
1\longrightarrow \mathrm{Sp}(2r, \mathbb{F}_{\ell})\longrightarrow \mathrm{GSp}(2r, \mathbb{F}_{\ell})\xrightarrow{\mathrm{mult}} \mathbb{F}_{\ell}^{*}\longrightarrow 1.
$$
For any $\gamma\in \mathbb{F}_{\ell}^{*}$, collect the `relevant' characteristic polynomials $f$ in the set
$$
M^{\gamma}_{r}:=\{f(T)=1+a_{1}T+\ldots +a_{2r-1}T^{2r-1}+\gamma^{r}T^{2r} \ \vert \ a_{i}\in \mathbb{F}_{\ell} ,\ a_{2r-i}=\gamma^{r-i}a_{i},\ 0\le i\le 2r \}.
$$
\end{definition}
We now give an estimate for the number of matrices with given characteristic polynomial $f(T)$. See \cite[Theorem 3.5]{chav} for a proof.
\begin{lemma}
\label{symplem}
   Fix $f(T)\in M^{\gamma}_{r}$. For $\ell>4$, we have
$$
(\ell -3)^{2r^{2}} \;\leq\; \#\{A\in \mathrm{GSp}(2r, \mathbb{F}_{\ell})^{\gamma} \ \vert \ f(T)=\mathrm{det}(1-TA) \} \;\leq\; (\ell+3)^{2r^{2}}\;.
$$
\end{lemma}
We may identify $M^{\gamma}_{r}$ with the points of the affine space $\mathbb{A}^{r}_{\mathbb{F}_{\ell}}$ with coordinates $(y_{1}, \ldots , y_{r})$, by sending a polynomial $f(T)=1+\sum_{i=1}^{2r-1}a_{i}T^{i}+\gamma^{r}T^{2r}$ to the tuple $(a_{1}, \ldots , a_{r})$.

Our goal is to obtain estimates for the proportion of characteristic polynomials that are \textit{not} coprime to a given $f(T)\in M_{r}^{\gamma}$. Let $W\subset \mathbb{A}^{r}_{\mathbb{F}_{\ell}}$ parameterise such polynomials. It is a hypersurface, given by the vanishing of $F(y_{1}, \ldots, y_{r})$, described as the resultant of a formal polynomial of the type 
$$g(T)\ =\ 1+\sum_{i=1}^{r}y_{i}T^{i}+\sum_{i=1}^{r-1}\gamma^{r-i}y_{i}T^{2r-i}+ \gamma^{r}T^{2r}$$
with $f(T)$ w.r.t.~$T$. The polynomial $F$ is of total degree at most $4r$ in the $y_{i}$. The number of its rational points, $\#W(\mathbb{F}_{\ell})$, gives the count we need. But, by \cite[pg 45]{bs}, we have
$
\#W(\mathbb{F}_{\ell})\leq 4r\ell^{r-1}.
$
Further, recalling the order formula for the symplectic group, we have
$$
\ell^{2r^{2}}(\ell-1)^{r}\ \leq\ \#\mathrm{Sp}(2r, \mathbb{F}_{\ell})\ =\ \ell^{r^{2}}\prod_{j=1}^{r}(\ell^{2j}-1)\ \leq\ \ell^{2r^{2}+r}\ .
$$
Therefore, combining with Lemma~\ref{symplem}, the proportion of matrices in $\mathrm{GSp}(2r, \mathbb{F}_{\ell})^{\gamma}$ with characteristic polynomial \textit{not} coprime to $f(T)$ is at most
$$
\frac{4r\ell^{r-1}\cdot (\ell+3)^{2r^{2}}}{\ell^{2r^{2}}(\ell-1)^{r}}\ =\ \frac{4r}{\ell}\left(1+\frac{1}{\ell-1}\right)^{r}\left(1+\frac{3}{\ell}\right)^{2r^{2}}\;,
$$
which is less than $1/4$, for $\ell>16e^{2}r^{2}$, where $e:=\mathrm{exp}(1)$. We summarise what we have shown in the following.
\begin{lemma}[Common eigenvalue]
\label{symplectic}
Let $r\in \mathbb{Z}_{>0}$ and let $\ell>4$ be a prime. Let $f(T)$ be the characteristic polynomial of a matrix in $\mathrm{GSp}(2r, \mathbb{F}_{\ell})^{\gamma}$ for some $\gamma\in \mathbb{F}_{\ell}^{*}$. Denote by $C\subset \mathrm{GSp}(2r, \mathbb{F}_\ell)$ the set of matrices with characteristic polynomial not coprime with $f(T)$. Then for $\ell>119r^{2}$,
$$
\frac{\#\left(C\cap \mathrm{GSp}(2r, \mathbb{F}_{\ell})^{\gamma}\right)}{\# \mathrm{Sp}(2r, \mathbb{F}_{\ell})} \;\leq\; 1/4\;.
$$

\end{lemma}

\section{$P_1(T)$ for smooth projective varieties}
\label{sec4}
This section proves the rest of our main theorems. \ref{red} details the reduction to the case of a surface and ~\ref{lef} delineates an algorithm to construct, with high probability, a Lefschetz pencil of hyperplane sections on a surface. In~\ref{eff}, we prove the effective gcd theorem, which forms the basis for the algorithms and the proofs of the Theorems~\ref{Theorem 1}, \ref{Theorem 3} and \ref{Theorem 2} in  Section~\ref{algsec}.
\subsection{Reduction to smooth projective surfaces}
\label{red}
In this subsection, we demonstrate a reduction of the problem of computing the characteristic polynomial of geometric Frobenius on the first ($\ell$-adic) \'{e}tale  cohomology of a smooth projective variety over a finite field $\mathbb{F}_{q}$ of fixed dimension $r>1$, to that of a smooth projective {\em surface}. This reduction is polynomial in the input data, namely the degree of the polynomials defining the variety and $\log q$. 
 
Let $X_{0}/\mathbb{F}_{q}$ be a smooth, projective, geometrically irreducible variety of dimension $n>1$ and degree $D>0$. We suppose that it is presented as a subvariety of $\mathbb{P}^{N}$, given by a homogeneous ideal $I$ generated by $m$ polynomials $f_{1}, \ldots , f_{m}$ of degree $\leq d$ for $d\in \mathbb{Z}_{>0}$. Denote by $X$ the base change to the algebraic closure. Let $\ell$ be a prime distinct from the characteristic of the base field. We recall the following.
 \begin{definition}
    Let $X$ be as above. A \textit{hyperplane section} of $X$ is a codimension 1 subvariety $Y\subset X$ obtained by intersecting $X$ with a hyperplane $H\subset \mathbb{P}^{N}$. A hyperplane $H$ is said to intersect $X$ \textit{transversally} at $x\in X$ if $T_{x}X\not\subset H$, i.e., $H$ does not contain the tangent space to $X$ at $x$. Equivalently, this translates to the condition that $X\cap H$ is smooth at $x$. In general, $H$ intersects $X$ \textit{transversally} if $H\cap X$ is a smooth, irreducible subvariety of codimension 1 of $X$.
\end{definition}
Denote by $(\mathbb{P}^{N})^{\vee}$ the dual projective space, parameterising hyperplanes in $\mathbb{P}^{N}$. We construct the \textit{dual variety} to $X$, denoted $\Check{X}\subset (\mathbb{P}^{N})^{\vee}$ as follows. Let 
$$
\Omega:=\{(x, H)\in X\times (\mathbb{P}^{N})^{\vee} \ \vert \ x\in H, \ T_{x}X\subset H \}.
$$
It is a closed subvariety of $X\times (\mathbb{P}^{N})^{\vee}$. We define $\check{X}$ to be the projection of $\Omega$ onto its second factor. In particular, $\check{X}$ parameterises those hyperplanes that \textit{do not} intersect transversally with $X$. We now state an effective version of Bertini's theorem, that ensures the availability of smooth hyperplane sections. The following is \cite[Theorem 1]{ballico}.
\begin{proposition}[Effective Bertini]
\label{bertini}
    Let $W\subset \mathbb{P}^{N}$ be a smooth, irreducible variety of dimension $n$ and degree $D$, defined over $\mathbb{F}_{q}$. Let $\mathbb{F}_{Q}/\mathbb{F}_{q}$ be an extension such that $Q>D(D-1)^{n}$. Then, there exists a hyperplane $H$ defined over $\mathbb{F}_{Q}$ that intersects transversally with $W$. 
\end{proposition}
In the proof of the above theorem, it is shown \cite[Lemma 1]{ballico} that $\check{W}$ is a variety of degree at most $D(D-1)^{N}\leq D^{N+1}$. The singular locus of $\check{W}$, denoted $\mathring{W}$ is a subvariety of $(\mathbb{P}^{N})^{\vee}$ of codimension at least $2$ and degree (by B\'ezout) at most $D^{(N+1)^{2}}$.
\begin{remark}
The existence of smooth hypersurface sections of sufficiently large degree is given by \cite{poonen}. However it is unavoidable to take field extensions for our algorithmic purposes (e.g., even to ensure the existence of a rational point), so the trade-off is immaterial.
\end{remark}
We now recall the following theorem, which is the key step in our reduction to surfaces. See \cite[\S8.5.5]{fu2011etale} for the proof of the more general theorem, of which this is a special case.
\begin{theorem}[Weak-Lefschetz]
\label{weaklefs}
Let $Y\hookrightarrow X$ be a smooth hyperplane section. Then the induced map
$$
\mathrm{H}^{1}(X, \mathbb{Q}_{\ell})\rightarrow \mathrm{H}^{1}(Y, \mathbb{Q}_{\ell}) 
$$
is an isomorphism if $n=\dim(X)>2$ and an injection if $n=2$.
\end{theorem}

With this setup, we notice that with an application of Bertini's theorem on the existence of smooth hyperplane sections, we can inductively reduce the dimension of $X$ by intersecting with a generic hyperplane in $\mathbb{P}^{N}$. In particular, there is a chain of smooth hyperplane sections $ Y:=Y_{2}\subset Y_{3}\subset ...\subset Y_{n-1}\subset X$, where $Y_{i}$ are smooth varieties of dimension $i$. Applying the weak-Lefschetz theorem, we get an isomorphism
$$
\mathrm{H}^{1}(X, \mathbb{Q}_{\ell})\rightarrow \mathrm{H}^{1}(Y, \mathbb{Q}_{\ell}),
$$
compatible with the action of the respective geometric Frobenii. 
Writing
$$
P_{1}(X/\mathbb{F}_{q}, T):=\mathrm{det}\left(1-TF_{q}^{\star}\ \vert \  \mathrm{H}^{1}(X, \mathbb{Q}_{\ell})\right)
$$
and assuming $Y$ is also defined over $\mathbb{F}_{q}$, we have
$P_{1}(X/\mathbb{F}_{q}, T)= P_{1}(Y/\mathbb{F}_{q}, T)$. So, it suffices to compute $P_{1}(Y/\mathbb{F}_{q},T)$ for $Y$ a smooth subvariety obtained from $X$ after intersection with $n-2$ hyperplanes in general position. 
\begin{remark}
    We note that the ideal defining $Y$ now is generated by the forms $f_{i}, L_{j}$ with $1\leq i\leq m$ and $1\leq j\leq n-2$, where the $L_{j}$ are linear forms representing the generic hyperplanes in $\mathbb{P}^{N}$ that we have intersected $X$ with, to obtain $Y$.
\end{remark}

\subsection{Lefschetz pencils on a surface}
\label{lef}
To study the zeta function of a surface, intuitively, one wants to break it up into those of curves, each parameterized by a single variable $t$, and then invoke the methods of Section \ref{sec2}. It is not so easy because Theorem \ref{weaklefs} does not give an isomorphism when $X$ is a surface, e.g., $\mathrm{H}^{1}(Y, \mathbb{Q}_{\ell})$ can be a larger group for a generic curve $Y$ lying on the surface $X$, which will make the zeta function of $Y$ `larger' than that of $X$, introducing errors called vanishing cycles (see Section \ref{sec-VC}).  

To explore these issues, in this subsection, we introduce the classic machinery of Lefschetz pencils and describe an algorithm to fibre a given smooth projective surface $X\subseteq \mathbb{P}^{N}$ of degree $D$ over the projective line so that the fibres are curves with singularities at worst being ordinary double points. We assume $X$ is given by $m$ homogeneous forms $f_{1},\ldots , f_{m}$, each of total degree $\leq d\in\mathbb{Z}_{>0}$. Denote by $(\mathbb{P}^{N})^{\vee}$, the dual projective space. 
\begin{definition}
 Let $X/\overline{\mathbb{F}}_{q}$ be as above. A \textit{Lefschetz pencil} on $X$ is a collection of hyperplanes $(H_{t})_{t\in \mathbb{P}^{1}}$ such that there exists a line $L\simeq \mathbb{P}^{1}\subset (\mathbb{P}^{N})^{\vee}$; for e.g., $(\lambda, \mu)\mapsto \lambda F=\mu G$, for linear forms $F,G$ on $\mathbb{P}^{N}$, satisfying the following conditions
 \begin{itemize}
     \item the \textit{axis}, of the pencil, $\Delta:=(F=0)\cap(G=0)$ in $\mathbb{P}^{N}$ intersects $X$ transversally, i.e., $X\cap \Delta$ is smooth of codimension 2,

     \item there is a dense open subset $U\subset \mathbb{P}^{1}$ on which the associated intersections $(\lambda, \mu)\rightarrow X\cap (\lambda F= \mu G)$ are smooth and geometrically irreducible for $(\lambda, \mu)\in U$; and have only an ordinary double point as singularity for the finitely many $(\lambda, \mu)\notin U$. 
 \end{itemize}
\end{definition}

It is a fundamental theorem that Lefschetz pencils exist on any smooth projective variety of dimension $\geq 2$, over an algebraically closed field (see \cite{katz1973pinceaux}). Over arbitrary fields, Lefschetz pencils exist, subject to a degree $\geq 3$ Veronese embedding.\footnote{this adds an overhead of only a polynomial in the degree $D$ of $X$.} We recall \cite[Theorem 3]{jannsensaito}.
\begin{proposition}
   There exists a nonempty open subscheme (after possibly passing to a degree $\geq 3$ Veronese embedding) in the Grassmannian of lines  $W_{X}\subset \mathrm{Gr}(1, (\mathbb{P}^{N})^{\vee})$ such that every $L\in W_{X}$ defines a Lefschetz pencil for $X$.
\end{proposition}
Algorithmically, to construct a Lefschetz pencil, we first take a field extension to ensure the existence of a transversal hyperplane section. We saw that the dual variety $\check{X}$ parameterises those hyperplanes that do not intersect transversally with $X$. Further, its singular locus $\mathring{X}$ parameterises those hyperplanes that intersect $X$ with singularity worse than a single ordinary double point. In other words, $\check{X}\setminus \mathring{X}$ consists of those hyperplanes $H$ such that $H\cap X$ has a single node (see \cite[Theorem 31.2]{milnelec} or \cite{katz1973pinceaux}). In light of Proposition~\ref{bertini}, it suffices to randomly take two linear forms $F,G\in (\mathbb{P}^{N})^{\vee}$. With high probability, they intersect transversally with $X$ and the line joining them in $(\mathbb{P}^{N})^{\vee}$ intersects $\check{X}$ at finitely many points and completely misses $\mathring{X}$.

\begin{algorithm}
\caption{Lefschetz pencil on a {\em surface}}
\label{alg:lef}
\begin{itemize}
    \item \textbf{Input:} A smooth projective {\em surface} $X_{0}/\mathbb{F}_{q}$ of degree $D$ presented as a system of homogeneous polynomials of degree $\leq d$ in the projective space $\mathbb{P}^{N}$. 
    \item \textbf{Pre-processing:} Replace $X$ with the degree 3 Veronese image of $X$ in $\mathbb{P}:=\mathbb{P}^{\binom{N+3}{3}-1}$.
  
    \item \textbf{Output:} Hyperplanes $F$ and $G$ in $\mathbb{P}$ such that the line $L$ through them in the dual $(\mathbb{P})^{\vee}$, is a Lefschetz pencil on $X$.
    
\end{itemize}
\begin{algorithmic}[1]

  \STATE Take a field extension $\mathbb{F}_{Q}/\mathbb{F}_{q}$ with degree bounded by a polynomial in $D$, such that smooth hyperplane sections exist as in Proposition~\ref{bertini}.

  \STATE Select two random linear forms $F$ and $G$ on $\mathbb{P}$, such that they intersect transversally with $X$ (this is possible by Proposition~\ref{bertini}).
  \STATE The line $L$ in $(\mathbb{P})^{\vee}$ through $F$ and $G$ is a Lefschetz pencil on $X$.
  
\end{algorithmic}
\end{algorithm}

\begin{lemma}
Algorithm~\ref{alg:lef} succeeds with probability at least $1-O(1/Q)$.
\end{lemma}
\begin{proof}
Indeed for $Q\gg D$, the locus of hyperplanes in $\mathbb{P}^{\vee}$ defined over $\mathbb{F}_{Q}$ that do not intersect transversally with $X$ is given by the dual variety $\check{X}$, which by the Lang-Weil estimates, can be avoided with probability $1-O(1/Q)$. Further, for two hyperplanes $H_{1}$ and $H_{2}$ that intersect transversally with $X$, the condition that they define a Lefschetz pencil on $X$ is equivalent to the condition that the line through the corresponding points in $\mathbb{P}^{\vee}$ does not intersect the singular locus $\mathring{X}$ of $\check{X}$. For two randomly chosen hyperplanes, this is also  ensured with probability greater than $1-O(1/Q)$, again by a Lang-Weil argument.

One checks that the output is correct by computing the finite subset $Z$ of `bad' hyperplanes (which is possible in poly-time) and verifying that the associated fibres are indeed nodal curves. The latter can be done by blowing up at a singular point and checking that the exceptional divisor intersects the transformed curve at two points, which has a polynomial-time algorithm.
\end{proof}

Blowing up $X$ along $X\cap \Delta$ gives a smooth projective surface $\tilde{X}$ and a morphism $\pi: \tilde{X}\rightarrow \mathbb{P}^{1}$ such that the fibre of a $\left[\lambda:\mu\right]\in \mathbb{P}^{1}$ is the curve $X\cap (\lambda F=\mu G)$. Algorithmically, the locus $\Delta\cap X$ may not all be defined over $\mathbb{F}_{q}$ and going to a field extension which contains all of the points therein may be expensive. Further, computing the blowup $\tilde{X}\rightarrow X$ and the morphism $\pi:\tilde{X}\rightarrow \mathbb{P}^{1}$ may also be exponential in the input data. Fortunately, we are able to leave $\pi$ implicit, i.e., the only knowledge we need is that the fibre of $u\in \mathbb{P}^{1}$ under $\pi$ is $H_{u}\cap X$, where $H_{u}$ is the hyperplane associated to $u$. We describe the required construction in Algorithm~\ref{alg:lef}. 

Now, consider the \'etale sheaf $R^{1}\pi_{\star}\mathbb{Q}_{\ell}$ on $\mathbb{P}^{1}$. It is locally constant of rank $2g$ on $U$, where $g$ is the genus of the generic fibre $\tilde{X}_{\eta}$ (which is a curve over the function field of $\mathbb{P}^{1}$), where $\eta\rightarrow \mathbb{P}^{1}$ is a geometric generic point. By the proper base-change theorem, for a point $u\in \mathbb{P}^{1}$, we have $(R^{1}\pi_{\star}\mathbb{Q}_{\ell})_{u}\simeq \mathrm{H}^{1}(\tilde{X}_{u}, \mathbb{Q}_{\ell})=\mathrm{H}^{1}(X\cap H_{u}, \mathbb{Q}_{\ell})$. Further, by \cite[Lemma 33.2]{milnelec}, we have $\mathrm{H}^{1}(\tilde{X}, \mathbb{Q}_{\ell})\simeq \mathrm{H}^{1}(X, \mathbb{Q}_{\ell})$.

We now establish bounds for the genus $g$ of the generic smooth fibre and for the critical locus which we call $Z:=\mathbb{P}^{1}\setminus U$. Pick $u\in U$, the fibre $X\cap H_{u}$ is a curve in $\mathbb{P}$ of degree $D=\mathrm{deg}(X)$. By the results of \cite{laz} (see also \cite[Theorem 3.3]{mack}), we have $$g\leq D^{2}-2D+1.$$ Further, the number of critical points, i.e., $\#Z$ is bounded by the size of $L\cap \check{X}$, which by the remark following Proposition~\ref{bertini} and B\'ezout's theorem, is at most $D^{N+1}$. Denote the Betti numbers $\beta_{i}:=\dim \mathrm{H}^{i}(X, \mathbb{Q}_{\ell})$. Clearly, we have\footnote{by Poincar\'e duality} $$\beta_{3}=\beta_{1} \leq 2g\leq 2D^{2}.$$
By \cite[V, Theorem 3.12]{milne1980etale}, we have $$\beta_{2}=\#Z+2\beta_{1}+2-4g\leq 2D^{N+1}.
$$
\subsection{An effective gcd theorem}
\label{eff}
 Now, let $X_{0}/\mathbb{F}_{q}$ be a smooth, projective, geometrically irreducible surface of degree $D>0$ obtained from good reduction of a smooth, projective surface $\mathcal{X}_{0}$ over a number field $K$ at a prime $\mathfrak{p}$. We assume that $X_{0}$ is presented as $X_{0}\subset \mathbb{P}^{N}$, given by a homogeneous ideal $I$ generated by $m$ polynomials $f_{1}, \ldots , f_{m}$ of degree $\leq d$ for $d\in \mathbb{Z}_{>0}$. Denote by $X$ the base change to the algebraic closure. Let $\ell$ be an odd prime, distinct from the characteristic, chosen according to Proposition~\ref{ellbound}. In this subsection, we prove an effective version of Deligne's `th\'eor\`eme du pgcd' \cite[Th\'eor\`eme 4.5.1]{Weilii}, that enables one to recover $P_{1}(X/\mathbb{F}_{q}, T)$ from the zeta function of hyperplane sections of $X$ (namely, simply by taking their gcd).
   
   Following Algorithm~\ref{alg:lef}, we may fibre $X$ as a Lefschetz pencil of hyperplane sections $\pi:\tilde{X}\rightarrow \mathbb{P}^{1}$ over $\mathbb{F}_{q}$ (after possibly replacing $\mathbb{F}_{q}$ by an extension of degree at most polynomial in $D$). Denote by $U\subset \mathbb{P}^{1}$ the open subscheme\footnote{write $U_{0}$ for the associated $\mathbb{F}_{q}$-scheme.} where the fibres are smooth, and $Z$ its complement. Let $g$ be the genus of the geometric generic fibre $X_{\eta}$.
 
Let  $u\in U(\mathbb{F}_{q})$. From the formalism of vanishing cycles and the so-called `hard-Lefschetz theorem' \cite[4.3.9]{Weilii}, we have the decomposition 
$$
\mathrm{H}^{1}(X_{u}, \mathbb{Q}_{\ell}) \,\simeq\, \mathrm{H}^{1}(X, \mathbb{Q}_{\ell})\oplus \mathcal{E}_{u}\,,
$$
where $X_{u}$ denotes the fibre of $\pi$ over $u$, and $\mathcal{E}_{u}$ is the space generated by the vanishing cycles in $\mathrm{H}^{1}(X_{u}, \mathbb{Q}_{\ell})$. In particular, we have that
$$
P_{1}(X_{u}/\mathbb{F}_{q}, T) \;=\; P_{1}(X/\mathbb{F}_{q}, T)\cdot P(\mathcal{E}_{u}/\mathbb{F}_{q},T)\;,
$$
where $P(\mathcal{E}_{u}/\mathbb{F}_{q}, T)$ denotes the characteristic polynomial of $F_{q}^{\star}$ acting on $\mathcal{E}_{u}$.
 
A theorem of Deligne (Theorem~\ref{gcd}) states that $P_{1}(X/\mathbb{F}_{q}, T)$ can be recovered from $P_{1}(X_{u_{i}}/\mathbb{F}_{q^{j}}, T)$ for $u_{i}\in U(\mathbb{F}_{q^{j}})$ over all extensions $\mathbb{F}_{q^{j}}$. We show that there is a `small' extension, and a small number of fibres over that extension to sample, to recover $P_{1}(X/\mathbb{F}_{q},T)$. 
 
Firstly, consider the representation $\rho:\pi_{1}(U_{0}, u)\rightarrow \mathrm{GL}(2r, \mathbb{Z}_{\ell})$ of the \'etale fundamental group of $U_{0}$ associated to the torsion-free lisse $\mathbb{Z}_{\ell}$-sheaf $\mathcal{E}_{\mathbb{Z}_{\ell}}\subset R^{1}\pi_{\star}\mathbb{Z}_{\ell}\vert_{U}$, of vanishing cycles. Denote by $\overline{\rho}:=\rho \ \vert \ \pi_{1}(U, u)$, the restriction to the geometric fundamental group. By \cite[5.10]{Weili}, we know that the Zariski-closure of the image of $\overline{\rho}\otimes \mathbb{Q}_{\ell}$ in $\mathrm{GL}(2r, \mathbb{Q}_{\ell})$ is the symplectic group $\mathrm{Sp}(2r, \mathbb{Q}_{\ell})$. Using methods of Hall \cite{hall}, we deduce that the mod-$\ell$ monodromy of the family, i.e., the image of $\overline{\rho}_{\ell}: \pi_{1}(U, u)\rightarrow \mathrm{GL}(2r, \mathbb{F}_{\ell})$  is the symplectic group $\mathrm{Sp}(2r, \mathbb{F}_{\ell})$.
 
Next, we note that for $u\in U(\mathbb{F}_{q^{j}})$ the `vanishing term' $P(\mathcal{E}_{u}/\mathbb{F}_{q^{j}}, T)$ is equidistributed (mod-$\ell$) in the family, \`a la Katz (see \cite[Theorem 4.1]{chav} or \cite[Theorem 9.7.13]{kasar}), so can be eliminated with high probability after two samplings. This is done by first moving to a large enough extension $\mathbb{F}_{Q}$ of $\mathbb{F}_{q}$ (to minimise the error term coming from the aforementioned equidistribution theorem) and sampling points uniformly randomly in $U(\mathbb{F}_{Q})$. Then the zeta functions of the associated fibres are computed and their gcd is taken. With high probability, this procedure gives $P_{1}(X/\mathbb{F}_{Q}, T)$, from which $P_{1}(X/\mathbb{F}_{q}, T)$ can be easily recovered using an analogue of Lemma~\ref{ked2}.

\begin{theorem}[Effective gcd] 
\label{effgcd}
There exists an extension $\mathbb{F}_{Q}/\mathbb{F}_{q}$, with degree $[\mathbb{F}_{Q}:\mathbb{F}_{q}]$  bounded by a polynomial in $D$, such that for any two distinct randomly chosen $u_{1}, u_{2}\in U(\mathbb{F}_{Q})$, we have with probability $>2/3$
$$
\mathrm{gcd}(P_{1}(X_{u_{1}}/\mathbb{F}_{Q}, T) , P_{1}(X_{u_{2}}/\mathbb{F}_{Q},T)) \;=\; P_{1}(X/\mathbb{F}_{Q}, T) \,.
$$

\end{theorem}\begin{proof}
    Let $\ell\in [(4D)^{4}\,, 2^{11}D^{N^{2}} ]$ be a prime distinct from $p$ such that the cohomology groups $\mathrm{H}^{i}(X, \mathbb{Z}_{\ell})$ are all torsion-free. This is possible by Proposition~\ref{ellbound}. 
    Consider the locally constant sheaf $R^{1}\pi_{\star}\mathbb{Z}_{\ell}\vert_{U}$ on $U$. It has as subsheaf, $\mathcal{E}_{\mathbb{Z}_{\ell}}$, the sheaf of $\mathbb{Z}_{\ell}$-vanishing cycles of rank (say) $2r$. Denote by $\rho: \pi_{1}(U_{0}, u)\rightarrow \mathrm{GL}(2r, \mathbb{Z}_{\ell})$ the associated $\ell$-adic representation and by $\overline{\rho}=\rho \vert \pi_{1}(U, u)$. Write $\rho_{\ell}$ and $\overline{\rho}_{\ell}$ respectively, for the mod-$\ell$ representations.
        
    By the hard-Lefschetz theorem, (Theorem~\ref{hardlef}) we have $2r=2g-\beta_{1}$ where $\beta_{1}$ is the first Betti number of $X$. By Theorem~\ref{bigm}, we know that the sheaf $\mathcal{E}_{\mathbb{Z}_{\ell}}$ has big mod-$\ell$ monodromy, i.e., $\mathrm{im}(\overline{\rho}_{\ell})=\mathrm{Sp}(2r, \mathbb{F}_{\ell})$. We seek to apply Theorem~\ref{katz} to this setup. Let $\mathbb{F}_{Q}/\mathbb{F}_{q}$ be an extension where $Q:= q^{w}$ and choose $u_{1}\in U(\mathbb{F}_{Q})$ randomly. We estimate the number of $v\in U(\mathbb{F}_{Q})$ such that $P(\mathcal{E}_{v}/\mathbb{F}_{Q},T)$ is coprime to $f(T):=P(\mathcal{E}_{u_{1}}/\mathbb{F}_{Q},T)$. Write $\overline{f}(T):=f(T) \ \mathrm{mod} \ \ell$.
    
     Denote by $C\subset \mathrm{GSp}(2r, \mathbb{F}_{\ell})$ the subset of matrices with characteristic polynomial not coprime to $\overline{f}(T)$. It is stable under conjugation by elements from $\mathrm{GSp}(2r, \mathbb{F}_{\ell})$. Applying Theorem~\ref{katz} to $C$, we get
$$
 \frac{\#\{v\in U(\mathbb{F}_{Q}) \ \vert \ \rho_{\ell}(F_{Q, v})\in C \} }{\#U(\mathbb{F}_{Q})} \;\leq\; \frac{\#(C\cap \mathrm{GSp}(2r, \mathbb{F}_{\ell})^{\gamma^{w}})}{\#\mathrm{Sp}(2r, \mathbb{F}_{\ell})}+ \vert \chi(U)\vert \frac{ \#\mathrm{GSp}(2r, \mathbb{F}_{\ell})\sqrt{q^{w}}}{\#U(\mathbb{F}_{Q})} \;.
$$
By Lemma~\ref{symplectic} (since $\ell> 119r^2$), the first summand on the RHS is $\leq 1/4$. From the calculation\footnote{see \cite[\href{https://stacks.math.columbia.edu/tag/03RR}{Tag 03RR}]{stacks-project}} of the \'{e}tale cohomology of $U$ (the projective line with $\#Z$ punctures), we deduce that $\vert \chi(U)\vert\leq\#Z\leq D^{N+1}$. For $q^{w}>2D^{N+1}$, we have
$$
\vert \chi(U)\vert \frac{ \#\mathrm{GSp}(2r, \mathbb{F}_{\ell})\sqrt{q^{w}}}{\#U(\mathbb{F}_{Q})} \;\leq\; D^{N+1}\ell^{2g^{2}+g+1}\frac{\sqrt{q^{w}}}{q^{w}-D^{N+1}} \;\leq\; 2D^{N+1}(2^{11}D^{N^{2}})^{4D^{2}}\frac{\sqrt{q^{w}}}{q^{w}/2} \;.
$$
In particular, if $Q=q^{w} > \Omega\left(D^{5N^{2}D^{4}}\right)$, we have
$$
\frac{\#\{v\in U(\mathbb{F}_{Q}) \ \vert \ \rho_{\ell}(F_{Q, v})\not\in C \} }{\#U(\mathbb{F}_{Q})} \;>\; 2/3 \,,
$$
which completes the proof.
    \end{proof}

\subsection{Algorithms for $P_{1}(T)$} 
\label{algsec}
Let $X_{0}\subset \mathbb{P}^{N}$  be a smooth projective variety of dimension $n>1$ and degree $D$ over $\mathbb{F}_{q}$, obtained via good reduction from $\mathcal{X}_{0}$; defined over a number field $K$, at a prime $\mathfrak{p}\subset \mathcal{O}_{K}$. An $\mathrm{AM}\cap \mathrm{coAM}$ protocol for certifying $P_{1}(X/ \mathbb{F}_{Q}, T)$ for any field extension $\mathbb{F}_{Q}/\mathbb{F}_{q}$ with $$Q> \Omega\left(D^{5N^{2}D^{4}}\right)$$ is presented in Algorithm~\ref{P1}.
\begin{algorithm}
\caption{Verifying $P_{1}(T)$ of a {\em variety}}
\label{P1}
\begin{itemize}
    \item \textbf{Input:} A smooth projective variety  $X_{0}/\mathbb{F}_{q}$ of dimension $n>1$ and degree $D$, presented as a system of $m$ homogeneous polynomials $f_{1}, \ldots, f_{m}$ of degree $\leq d$ in the projective space $\mathbb{P}^{N}$.
    \item \textbf{Pre-processing:}
   We first move to a field extension $\mathbb{F}_{Q}/\mathbb{F}_{q}$
    that affords enough smooth hyperplane sections as in Proposition~\ref{bertini} and satisfies the bound of Theorem~\ref{effgcd}. We may reduce to a surface $Y$ by intersecting $X$ with $n-2$ generic hyperplanes. Next, $Y$ is fibred as a Lefschetz pencil of hyperplane sections following Algorithm~\ref{alg:lef}. Denote by $U\subset \mathbb{P}^{1}$ the open subscheme parameterising the smooth fibres, and $Z:=\mathbb{P}^{1}\setminus U$ the finitely many singular ones.
    
    \item \textbf{Conditions:} Merlin provides a candidate $P(T)$ for $P_{1}(X/\mathbb{F}_{Q},T)$ and Arthur engages in a protocol with Merlin to determine the veracity of the claim.
\end{itemize}
\begin{algorithmic}[1]
  
  \STATE \textbf{Arthur:} Pick randomly distinct $u_{i}\in U(\mathbb{F}_{Q})$, for $1\leq i\leq 2$  following Theorem~\ref{effgcd}. 

  \STATE \textbf{Merlin:} Provide $P_{1}(Y_{{u}_{i}}/\mathbb{F}_{Q}, T)$, for $1\leq i\leq 2$.

  \STATE \textbf{Arthur:} Verify that the $P_{1}(Y_{{u}_{i}}/\mathbb{F}_{Q}, T)$ are as claimed by calling Algorithm~\ref{algzetacurve}. Compute their greatest common divisor $G(T)$, using e.g., Euclid's algorithm. Accept iff $G(T)=P(T)$.
  \end{algorithmic}
\end{algorithm}

\begin{reptheorem}{Theorem 1}
    Given $Q_1(T)\in \mathbb{Z}[T]$, deciding whether $Q_1(T)=P_{1}(X/\mathbb{F}_{q}, T)$, for $X_{0}$ given as above, is in $\mathrm{AM}\cap \mathrm{coAM}$.
\end{reptheorem}
\begin{proof}
    Let $X_{0}\subset \mathbb{P}^{N}$ be a smooth projective variety of dimension $n>1$ and degree $D$, over the field $\mathbb{F}_{q}$ given by homogeneous forms $f_{1}, \ldots, f_{m}$, each of total degree $\leq d\in \mathbb{Z}_{>0}$. For any extension $\mathbb{F}_{Q}/\mathbb{F}_{q}$ such that $Q> \Omega\left(D^{5N^{2}D^{4}}\right)$ of poly-bounded degree, we may verify $P_{1}(X/\mathbb{F}_{Q}, T)$ using Algorithm~\ref{P1}. Now, choosing two field extensions $\mathbb{F}_{Q_{1}}/\mathbb{F}_{q}$ and $\mathbb{F}_{Q_{2}}/\mathbb{F}_{q}$ with size greater than $ \Omega\left(D^{5N^{2}D^{4}}\right)$ according to Lemma~\ref{ked2}, we can recover and hence certify $P_{1}(X/ \mathbb{F}_{q}, T)$ as well.
\end{proof}

For Theorem~\ref{Theorem 3}, we recall a theorem of Kedlaya \cite[Theorem 1]{kedlaya2006quantum} that enables efficient quantum computation of the zeta function of a curve.

\begin{theorem}[Kedlaya]
\label{thm-quantum}
    Let $C\subset \mathbb{P}^{N}$ be a smooth projective curve over $\mathbb{F}_{q}$, of degree $D$. Then, there exists a quantum algorithm that computes $P_{1}(C/\mathbb{F}_{q}, T)$ in time polynomial in $D\log q$.
\end{theorem}

\begin{reptheorem}{Theorem 3}
    There exists a quantum algorithm that computes $P_{1}(X/\mathbb{F}_{q}, T)$ in time polynomial in $D \log q$, for any $X_{0}$ as given above.
\end{reptheorem}

\begin{proof}
 Similarly to Algorithm~\ref{P1}, we begin by reducing to the case of a surface $Y$ (obtained via successive hyperplane sections of $X$) and fibring as a Lefschetz pencil of hyperplane sections. We then move to a large enough field extension $\mathbb{F}_{Q}/\mathbb{F}_{q}$ as before, and sample $u_{i}\in U(\mathbb{F}_{Q})$ uniformly randomly for $i\in \{1,2\}$. Now, using Theorem~\ref{thm-quantum}, we may compute $P_{1}(Y_{u_{i}}/\mathbb{F}_{Q}, T)$ of the curves $Y_{u_{i}}$ and take their gcd. With probability $>2/3$, the result is $P_{1}(Y/\mathbb{F}_{Q}, T)=P_{1}(X/\mathbb{F}_{Q}, T)$. We use the technique of Lemma~\ref{ked2} to recover the characteristic polynomial $P_{1}(X/\mathbb{F}_{q}, T)$ of the base Frobenius  as well.

\end{proof}

We now recall the following result to compute the zeta function of a smooth curve of fixed degree. 
\begin{theorem}[Pila, Huang-Ierardi]
\label{pila}
Let $C\subset \mathbb{P}^{N}$ be a smooth projective curve over $\mathbb{F}_{q}$, of fixed degree $D$. Then, there exists an algorithm that computes $P_{1}(C/\mathbb{F}_{q}, T)$ in time $O((\log q)^{\Delta})$, where $\Delta$ is independent of $q$ and polynomial in $D$.
\end{theorem}
\begin{proof}
    Move to a plane nodal model $C'$ of $C$ via Lemma~\ref{planemodel} and apply  \cite[Theorem 1.1]{huaier}.
\end{proof}

\begin{reptheorem}{Theorem 2}
There exists a randomised algorithm that, given $X_{0}$ as above of fixed degree $D$, computes $P_{1}(X/\mathbb{F}_{q}, T)$ in time polynomial in $\log q$.     
\end{reptheorem}

\begin{proof}
Similar to the proof of Theorem~\ref{Theorem 3}, use the algorithm of Huang-Ierardi from Theorem~\ref{pila} to compute $P_{1}(Y_{u_{i}}/\mathbb{F}_{Q}, T)$ and, then take their gcd. Use Lemma~\ref{ked2} to recover the characteristic polynomial of the base Frobenius.

\end{proof}

\section{Conclusion}\label{sec6}
We have presented randomised methods to efficiently compute and certify the characteristic polynomial of the geometric Frobenius on the {\em first} $\ell$-adic \'etale cohomology of smooth varieties. The immediate question is for higher cohomologies: to begin with, how do we compute $P_{n}(T)$ for a smooth projective {\em variety} of dimension $n>1$ over $\mathbb{F}_{q}$ in time polynomial in $\log q$? In another direction (for variable $D\log q$), one may ask for \textit{deterministic} verification, i.e., an $\mathrm{NP}\cap\mathrm{coNP}$ protocol for $P_{1}(T)$ and more generally for $P_{i}(T)$. 


\section*{Acknowledgements}
 N.S. thanks the funding support from DST-SERB (CRG/2020/45 + JCB/2022/57) and N. Rama Rao Chair.  M.V. is supported by a C3iHub research fellowship. We thank Chris Hall, Donu Arapura and Jeff Achter for email conversations and Hyuk Jun Kweon for pointing out the work \cite{kweon}, which led to Proposition~\ref{ellbound}. We thank Vasudevan Srinivas, Kiran Kedlaya and Partha Mukhopadhyay for enthusiastic discussions about the related problems. We thank the many researchers who provided valuable feedback by attending our talks and reading the draft.

\bibliographystyle{alpha}
\bibliography{L1.bib}

\end{document}